\pdfoutput=1

\documentclass[preprints,article,accept,moreauthors,pdftex,10pt,a4paper]{mdpi} 

\preto{\abstractkeywords}{\nolinenumbers}

\firstpage{1} 
\pubvolume{xx}
\issuenum{1}
\articlenumber{5}
\pubyear{2018}
\copyrightyear{2018}
\history{Received: date; Accepted: date; Published: date}

\usepackage{xspace}
\usepackage{overpic}
\usepackage{gensymb}
\usepackage[caption=false]{subfig}
\usepackage{bookmark}
\usetikzlibrary{
	matrix,
	shapes,
	arrows,
	positioning,
	chains,
	fit,
	backgrounds,
	calc}
\tikzstyle{startstop} = [rectangle, rounded corners, minimum width=3cm, minimum height=1cm,text centered, draw=black, fill=black!10]
\tikzstyle{io} = [trapezium, trapezium left angle=70, trapezium right angle=110, minimum width=3cm, minimum height=1cm, text centered, draw=black, fill=black!10]
\tikzstyle{process} = [rectangle, minimum width=4cm, minimum height=1cm, text centered, text width=4cm, draw=black, fill=black!10]
\tikzstyle{decision} = [diamond, minimum width=3cm, minimum height=1cm, text centered, text width = 3cm, draw=black, fill=black!10]
\tikzstyle{arrow} = [thick, dashed,->,>=stealth]
\newcommand{\ie}{\textit{i.e.}\xspace}
\newcommand{\eg}{\textit{e.g.}\xspace}
\newcommand{\hbpgrant}{Specific Grant Agreements No. 785907 (HBP SGA2) and No. 720270 (HBP SGA1)\xspace}

\Title{Analysis and Model of Cortical Slow Waves Acquired with Optical Techniques}

\Author{Marco Celotto $^{1,6}$\orcidG{}, Chiara De Luca $^{1,5,7}$\orcidA{}, Paolo Muratore $^{1,8}$\orcidH{}
		Francesco Resta $^{2}$\orcidB{}, Anna Letizia Allegra Mascaro $^{2,3}$\orcidC{}, Francesco Saverio Pavone $^{2,4}$\orcidD{},
        Giulia De Bonis $^{5,}$*\orcidE{} and Pier Stanislao Paolucci $^{5}$\orcidF{}}
\AuthorNames{Marco Celotto, Chiara De Luca, Paolo Muratore, Francesco Resta, Anna Letizia Allegra Mascaro, Francesco Pavone, Giulia De Bonis and Pier Stanislao Paolucci}

\address{%
$^{1}$ \quad Department of Physics,``Sapienza'' University of Rome, 00185 Rome, Italy; celotto.1650378@studenti.uniroma1.it (M.C.); deluca.1665541@studenti.uniroma1.it (C.D.L.); muratore.1668707@studenti.uniroma1.it (P.M.) \\
$^{2}$ \quad LENS, University of Florence, 50019 Florence, Italy; resta@lens.unifi.it (F.R.); allegra@lens.unifi.it (A.L.A.M.); pavone@lens.unifi.it (F.S.P.)\\
$^{3}$ \quad Istituto di Neuroscienze, CNR, 56124 Pisa, Italy\\
$^{4}$ \quad Department of Physics, University of Florence, 50019 Florence, Italy\\
$^{5}$ \quad INFN, 00185 Rome, Italy; giulia.debonis@roma1.infn.it; pier.paolucci@roma1.infn.it \\
$^{6}$ \quad IIT - Neural Computation Lab, CNCS@UniTn, 38068 Rovereto, Italy; marco.celotto@iit.it \\
$^{7}$ \quad PhD Program in Behavioural Neuroscience,``Sapienza'' University of Rome, 00185 Rome, Italy \\ 
$^{8}$ \quad PhD Program in Cognitive Neuroscience, SISSA, 34136 Trieste, Italy; pmurator@sissa.it
}

\corres{Correspondence: giulia.debonis@roma1.infn.it}

\abstract{Slow waves (SWs) are spatio-temporal patterns of cortical activity that occur both during natural sleep and anesthesia and are preserved across species. 
Even though electrophysiological recordings have been largely used to characterize brain states, they are limited in the spatial resolution and cannot target specific neuronal population. Recently, large-scale optical imaging techniques coupled with functional indicators overcame these restrictions, and new pipelines of analysis and novel approaches of SWs modelling are needed to extract relevant features of the spatio-temporal dynamics of SWs from these highly spatially resolved data-sets.
Here we combined wide-field fluorescence microscopy and a transgenic mouse model expressing a calcium indicator (GCaMP6f) in excitatory neurons to study SW propagation over the meso-scale under ketamine anesthesia. 
We developed a versatile analysis pipeline to identify and quantify the spatio-temporal propagation of the SWs. 
Moreover, we designed a computational simulator based on a simple theoretical model, which takes into account the statistics of neuronal activity, the response of fluorescence proteins and the slow waves dynamics. The simulator was capable of synthesizing artificial signals that could reliably reproduce several features of the SWs observed \textit{in vivo}, thus enabling a calibration tool for the analysis pipeline.
Comparison of experimental and simulated data shows the robustness of the analysis tools and its potential to uncover mechanistic insights of the Slow Wave Activity (SWA).}


\begin{document}


\section{Introduction} 
\label{intro}

The phenomenon of slow cortical waves (\textit {delta waves}) is a regime of brain activity that is observed in all mammals in a state of deep sleep or under anaesthesia\cite{Sanchez2017:Shaping-the-Default-Activity, Nghiem2018:Cholinergic}. It is characterized by a large-scale collective activation of groups of neurons, with a characteristic undulatory space-time pattern. Slow waves appear to propagate throughout the cortex modulating the spiking frequency of the underlying neurons populations. More specifically, as a slow wave passes, the involved neurons transit from a state of low spiking activity (\textit{down-state}) to a more intense one (\textit{up-state}).
Since the first electroencephalogram (EEG) observations in the 1930s, many experimental studies have been conducted on the large-scale activity of neuronal populations. These studies have permitted to observe the brain moving between various states of activity, according to the cognitive state of the subject. The awake activity is distinguished from sleep, which is in turn subdivided into the Rapid Eye Movements (REM) stage (when the brain activity is comparable with the awake state) and in 3 non-REM (NREM) stages \cite{Malhotra2013:SleepStages}. NREM-3 identifies the deepest sleep state, characterized by \emph{delta waves} that express in the lowest part of the frequency spectrum ($\left[0.5, 4 \right]\ Hz $); the cortical activity during NREM-3 is emulated by deep anesthesia states.

%

In recent years, the comprehension of both the dynamics and the functional role of delta waves significantly improved. Current studies outline two complementary roles of slow waves, respectively implemented by the down-states and the up-states of neuronal populations \cite{Vyazovskiy2013:Sleep-and-the-single-neuron, Killgore2010:Effects-of-sleep-deprivation, Buszaki2016:NetworkHomeostasis}. According to these studies, the down-states would help neurons to rest, favoring their prophylactic maintenance; in other words, they would guarantee the restorative function of sleep \cite{Vyazovskiy2013:Sleep-and-the-single-neuron}. On the other hand, the up-states would have a fundamental role in learning: the induction of spikes with appropriate temporal delays between pre- and post-synaptic neurons would trigger the mechanism of synaptic plasticity, enhancing and weakening the synaptic efficacies in order to encode the memories acquired during wakefulness \cite{Buszaki2016:NetworkHomeostasis, Paolucci2018:Sleep-like-slow-oscillations}. Indeed, several studies showed the negative consequences of sleep deprivation on many cognitive functions, such as memories consolidation and psychomotor vigilance \cite{Killgore2010:Effects-of-sleep-deprivation}.
	
Up to the present, electrophysiological recordings such as electroencephalography (EEG) and electrocorticography (ECoG) have been the standard methods for collecting slow wave data in humans. When combined with more invasive techniques like extracellular and intracellular recording both \textit{in vivo} and \textit{ex vivo}, animal models provide many entry points toward the investigation of neuronal functionality  \cite{Sanchez-Vives2000:Cellular-and-network-rythmic, Capone2017:Slow-Waves-Cortical-Slices, Steriade1993:Novel-slow}.
Recently, cortex-wide mesoscopic optical imaging, combined with fluorescent indicators of activity, provided new insight on the spatio-temporal propagation pattern of the brain activity. Few studies took advantage of Voltage Sesitive Dyes (VSDs) to unravel the dynamics of cortical traveling waves across different anesthetic brain states, during slow wave sleep and during resting wakefulness in mice (\cite{Mohajerani2013:SpontaneousCorticalActivity, Greenberg2017:New-waves}). Genetically encoded proteins are much less invasive than synthetic dyes, allow repeated measures on the same sample, and can be targeted to specific cell populations. The combination of encoded calcium and voltage indicators (GECIs and GEVIs, respectively) with wide-field microscopy is a unique approach for examining the neuronal activity in anesthetized and awakening mice at the mesoscale \cite{Akemann2012:Imaging-neural-circuit, Konnerth2012:Imaging-Calcium, Chen2013:Ultrasensitive-fluorescent, Shimaoka2017:State-dependent-modulation, Scott2014:Voltage-imaging, Fagerholm2016:Cortical-entropy, Wright2017:Functional-connectivity, Vanni2017:Mesoscale-Mapping, Xie2016:Resolution-of-High-FrequencyMesoscaleIntracorticalMaps}).
In detail, GECIs enable to visualize fluctuations in calcium concentration, which is an indirect reporter of neuronal spiking activity \cite{Konnerth2012:Imaging-Calcium}. Compared to GEVIs, recently developed GECIs like GCaMP6 are characterized by a very high sensitivity, and within the GCaMP family, GCaMP6f (f = fast) represents the best compromise between sensitivity and kinetic speed in response to the presence of calcium (\cite{Chen2013:Ultrasensitive-fluorescent}).
%
By allowing mapping of spontaneous network activity over a broad range of frequencies (including the delta band), GECIs recently proved to be a powerful tool to dissect the spatio-temporal features of slow wave activity \cite{Wright2017:Functional-connectivity, Vanni2017:Mesoscale-Mapping}. Despite the lower temporal resolution compared to electrophysiological methods, this approach increased both the spatial resolution and the specificity of the recorded neuronal population. 

Here we took advantage of large scale fluorescence microscopy coupled to transgenic mice expressing GECIs in excitatory neurons. 
The description we adopt is that the passage of a slow wave (in the delta band) over a neuronal population causes the spiking frequency of each involved neuron to move from $\mu_{\mathrm {down}}$ (in the Down state) to $\mu_{\mathrm {up}}$ (in the Up state).
Indeed, in this picture, the activity of the cortex during deep sleep is characterized by bi-stability, with neurons oscillating between two regimes, or modes, and bi-modality modulated by the passage slow waves, that induce a coherent state transition in the invested cell assemblies. The timestamp of the transition has to be identified to quantify the slow wave dynamics across the cortex surface, and we assume (as we will discuss in details below) that the dynamics of the fluorescence response function is fast enough to identify, at least, the presence of down- to up-state wavefronts.
Rooted in this framework, and based on the time series extracted from the wide-field images on one cortical hemisphere of anesthetized mice (described in Section \ref{subsec:Methods-DAQ}), we (i) developed a pipeline of data analysis 
that can evaluate the main spatio-temporal features of wave propagation (Section \ref{subsec:DataAnalysisPipeline}); (ii) built and validated a simple model of neuronal population activity (the Toy Model) that could simulate the propagation of cortical waves, to be employed as a calibration tool for the data analysis pipeline (Section \ref{subsec:ToyModel}). Section \ref{sec:results} illustrates the results obtained on the propagation of neuronal activity throughout the cortex; Section \ref{sec:Discussion} is for discussion and conclusions, with some remarks on future developments of the methods here presented and on further studies enabled by this approach. 


\section{Materials and Methods}
\label{sec:Materials-and-Methods}
Within this section, we describe the experimental procedures followed to collect the experimental data; then, we focus on the analysis pipeline and on the Toy Model.

	
\subsection{Experimental procedures and data processing}
\label{subsec:Methods-DAQ}

Experimental data, acquired from mice, have been provided by LENS (European Laboratory for Non-Linear Spectroscopy\footnote{LENS Home Page, \url{http://www.lens.unifi.it/index.php} (accessed on Nov. 2019).}). All procedures involving mice were performed in accordance with the rules of the Italian Minister of Health (Protocol Number 183/2016-PR). 
Mice were housed in clear plastic cages under a 12 h light/dark cycle and were given \textit{ad libitum} access to water and food. 

Mouse Model: We used a transgenic mouse line from Jackson Laboratories (Bar Harbor, Maine USA), the C57BL/6J-Tg(Thy1GCaMP6f)GP5.17Dkim/J (referred to as GCaMP6f mice\footnote{For more details, see The Jackson Laboratory, Thy1-GCaMP6f, \url{https://www.jax.org/strain/025393} (accessed on Nov. 2019).}). In this mouse model, the calcium indicator GCaMP6f is selectively expressed in excitatory neurons \cite{Dana2014:Thy1-GCaMP6}. This fluorescent protein is ultra-sensitive to calcium ions concentration \cite{Chen2013:Ultrasensitive-fluorescent, Dana2014:Thy1-GCaMP6}, and the fluorescence signal associated with fluctuations in calcium concentration is eventually correlated with the firing rate of neuronal populations underlying each pixel \cite{Konnerth2012:Imaging-Calcium, Yasuda2004:Imaging-calcium-concentration}.

Surgery: Mice are anaesthetized with a mix of Ketamine and Xylazine in doses of $100\ mg/kg$ and $10\ mg/kg$ respectively. Ketamine is an antagonist of the postsynaptic NMDA receptors \cite{Irifune1992:Ketamine}, largely used in literature to study anesthesia-related slow waves \cite{Steriade1993:Slow-oscillation, Chauvette2011:Properties-of-slow-oscillation}. 
To obtain optical access to neuronal activity over the right hemisphere, the skin over the skull and the periosteum were removed, after applying the local anesthetic lidocaine ($20\ mg/mL$). Wide-field imaging was performed right after the surgical procedure.  

Wide-field Fluorescence Microscopy: For imaging of GCaMP6f fluorescence, a $505\ nm$ LED light (M505L3 Thorlabs, New Jersey, United States) was deflected by a dichroic filter (DC FF 495-DI02 Semrock, Rochester, New York USA) on the objective (2.5x EC Plan Neofluar, NA 0.085, Carl Zeiss Microscopy, Oberkochen, Germany). A 3D motorized platform (M-229 for the $xy$ plane, M-126 for the $z$-axis movement; Physik Instrumente, Karlsruhe, Germany) allowed sample displacement. The fluorescence signal was selected by a band-pass filter (525/50 Semrock, Rochester, New York USA) and collected on the sensor of a high-speed complementary metal-oxide semiconductor (CMOS) camera (Orca Flash 4.0 Hamamatsu Photonics, NJ, USA). The fluorescence signal recorded in wide-field imaging originates from multiple cells in the cortical volume. 

%
The dataset consists of a series of $100 \times 100$ pixels \textit{.tif} images spaced out by a time step of $40\ ms$ (sampling frequency of $25\ Hz$). Each image covers a $25\ mm^2$ ($5 \ mm \times 5\ mm$) area and offers a perspective from above on the brain cortex of two transgenic Thy1-GCaMP6f mice (hereafter, referred to as \textit{Keta1} and \textit{Keta2}).
The total collection time for each mouse is about 5 minutes (8 sets of $1000$ frames, each set corresponding to an observation period of $ 40\ ms \times 1000 = 40\ s$). Modern high-density electrode arrays for ECoG recordings can interrogate up to 32 sites per hemisphere in mice and can reach a sampling frequency of $25\ kHz$ when the signal from each channel is digitized \citep{Konerding2018:New-thin-film,Pazzini2018:UltraCompactSystem,DeBonis2019:SWAP}. Compared to this, the resolution of wide field optical microscopy is much worse from a temporal point of view ($ 25\ Hz $ for the optical technique, up to $ 25\ kHz $ for the electrode grid), but spatially higher (for the optical technique, each pixel is $ 50\ \mu m \times 50\ \mu m$, to compare with the $ 500 \ \mu m $ electrode spacing for the grid that corresponds to a worsening of the spatial inspection by a factor 100\footnote{The spacing of the electrodes can be smaller than $ 500 \ \mu m $ for ultra-thin flexible arrays \citep{Khodagholy2014:Neurogrid}, approaching the spatial resolution of imaging recordings.}).



\begin{figure}[h]
	\centering
	\includegraphics[width=0.48\textwidth,trim={0 3.5cm 0 3.1cm},clip]{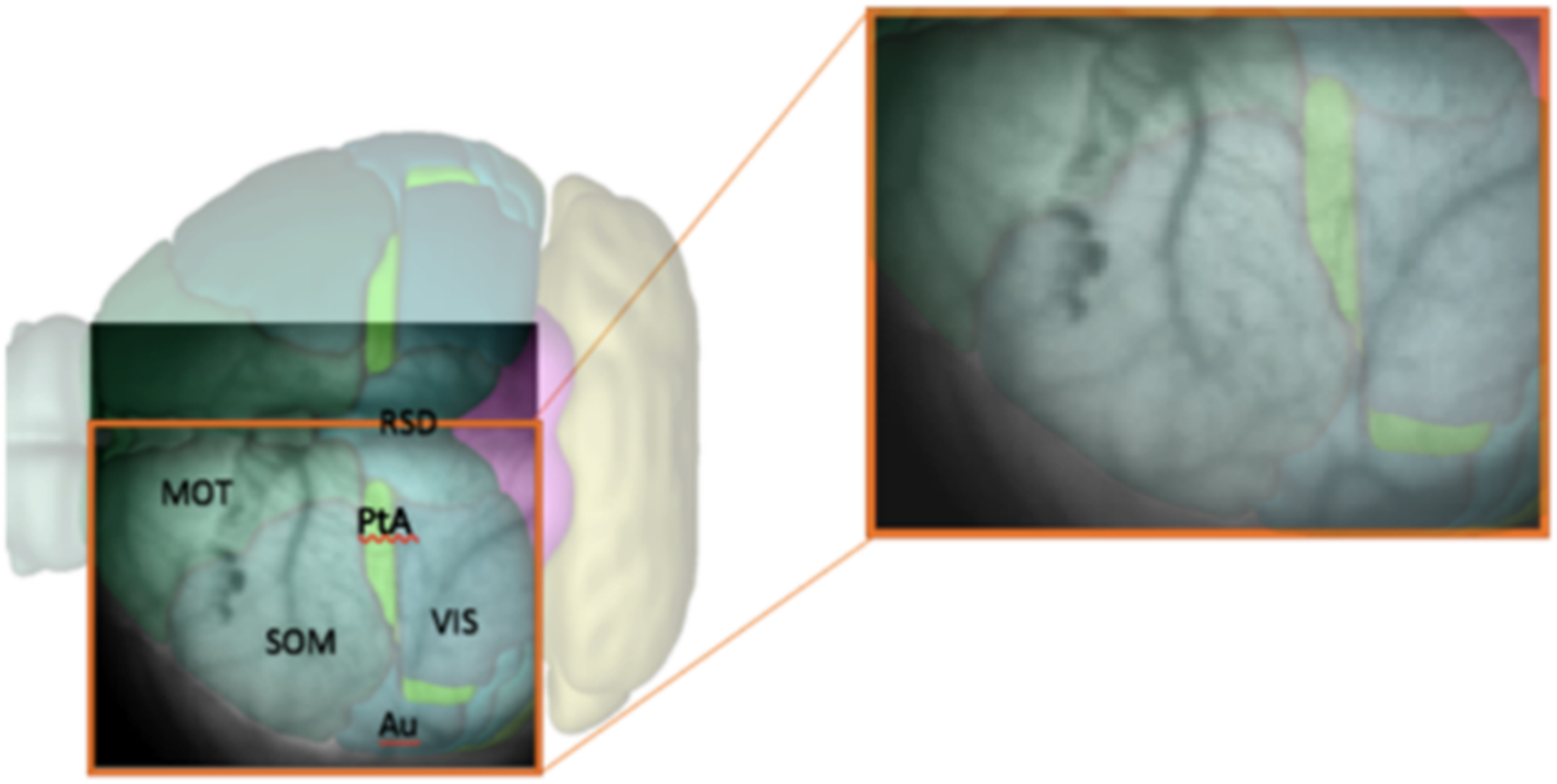}
	\caption{\small Representation of the mouse cerebral cortex  with overlapping names of the different areas: MOT = motor; SOM = somatosensory; Au = auditory; VIS = visual; PtA = associative parietal; RSD = retrosplenial dysgranular. The zoom shows which part of the cortex is visible in the raw data at our disposal, after selecting the left hemisphere and before the removal of the black background.}
	\label{fig:aree_cervello}
\end{figure}

In every analyzed image the left cerebral hemisphere of the anaesthetized mouse is clearly visible, while the incomplete parts of the right hemisphere have been cut off, if present, at the beginning of the analysis pipeline. 
In addition, the black background appearing in the image has been masked since not informative.
After this first processing steps, the $100 \times 100$ pixels raw images are reduced in size.
In Figure \ref{fig:aree_cervello}, a sample image is superimposed on the atlas of a mouse's brain cortex, in order to tag different cortical areas in the data.
	
\subsection{Data Analysis Pipeline}
\label{subsec:DataAnalysisPipeline}

The data analysis pipeline (named CaImanSWAP\footnote{\textit{CaImanSWAP} stands for Calcium Imaging Slow Wave Analysis Pipeline; SWAP is the name given to the software tools developed for the study of electrophysiological recordings \cite{DeBonis2019:SWAP}.}) has been coded in Python, with the employ the \emph{scikit-image} package\footnote{Image Processing in Python, \url{https://scikit-image.org} (accessed on Nov. 2019).} for the processing of the raw data.

The pipeline consists of three main sequences. The first concerns image initialization and signal cleaning, and it is further subdivided into three parts:
	
\begin{itemize}
	\item \textbf{Image initialization}. Each \textit{.tif} image 
	is imported and translated in a matrix of floats. Matrices are then assembled so that a collection of images belonging to the same recording set is stored in a 3D-array: 2 dimensions are spatial (representing the cortex surface) while the third is time (that is quantized in $40\ ms$ steps). Second, data are manipulated in order to keep informative content only, as already illustrated in Figure \ref{fig:aree_cervello}. Matrices are cut in the spatial dimensions, isolating the portion of interest of the dataset (\eg eliminating parts of the right hemisphere). Afterwards, in order to remove the non-informative black background, we make use of a masking process. The edges of the mask are obtained with the \textit{find\_contours} function (\textit{skimage.measure} Python package)  which adopts the \textit{marching squares} algorithm; each pixel outside the mask is forced to take the \textit{NaN} value (\emph{Not A Number}). 
	Since images in the same recording set share the same perspective on the cortex, the cropping process is applied \textit{en bloc} to each dataset.
	The cropping process (cutting and masking) is the step of the analysis procedure that cannot be fully automatized at this stage of software development, because it depends on the quality of the image set under study. 
	\item \textbf{Background subtraction and spatial smoothing}. Once the informative parts of the data are isolated, the pipeline proceeds with signal cleaning. The constant background is estimated for each pixel as the mean of the signal computed on the whole image set; it is then subtracted from images, pixel by pixel. 
	Fluctuations are further reduced with a spatial smoothing: pixels are assembled in $2\times2$ blocks (\textit{macro-pixels}); at each time, the value of a macro-pixel is the mean signal of the $4$ pixels belonging to it\footnote{From now on, \textit{macro-pixel} is meant each time \textit{pixel} is mentioned.}. This step may be unnecessary given the cleanness and regularity of the datasets we have used to test and develop the analysis pipeline, but our goal when devising the analysis procedure was to arrange a set of methods that could have been applied to any dataset, including noisy cases. The current implementation with macro-pixels corresponds to a factor $1/4$ of reduction in the number of inspected channels, with an equivalent grid step of $100\ \mu s$ for the pixel array, smaller than or comparable with the size currently accessible with surface multi-electrode arrays (MEAs), keeping this analysis still competitive in term of spatial resolution. 
	\item \textbf{Spectrum analysis and time smoothing}. In order to identify the dominant frequency band in each dataset, a real Fourier transform (RFT) is computed for each pixel. A unique spectrum for each dataset is then obtained as the mean of single pixels' RFT. Given the mean spectrum, the frequency band of maximum intensity is identified and the signal is further cleaned by applying a $6\degree$ order band-pass Butterworth filter (provided by the \textit{scipy} Python package\footnote{Scientific Python, \url{https://www.scipy.org} (accessed on Nov. 2019).}).

	
\end{itemize}
	
At the end of these preliminary steps, we obtain a 3D-array containing only the informative and cleaned signal. 
Subsequently, for normalization purposes, the signal for each pixel is divided by its maximum value; the unit is therefore $\frac{F - F_{back}}{F_{max} - F_{back}} \equiv \frac{\mathcal{F}}{\mathcal{F}_{max}}$ . 
	
\vspace{0.1cm}
	
The second part of the analysis consists of finding the set of upward transition times for each pixel, and of slicing the collection of transition times into waves. As already introduced, a wave is a coherence signal that propagates along the cortex surface and solicits the cell assemblies, having the effect of inducing a transition from Down to Up state in the involved neuronal populations. The frequency of the phenomenon is in the delta band, meaning that between 1 and 4 travelling signals (waves) can be observed in a second of recordings. 
The most common definitions of the timestamp of the upward transition are: (i) the onset of the transition; (ii) the crossing of a threshold that separates Up and Down states. This second option is not suitable in the case of calcium imaging, because of the timing of the transfer function\footnote{With the term \textit{transfer function} we refer to the curve shown in \citep{Chen2013:Ultrasensitive-fluorescent}, Figure 1.b, in analogy with the transfer function of a filter: indeed, the effect of obtaining the luminosity signal as the convolution of the neuronal spiking activity with the response function illustrated in \citep{Chen2013:Ultrasensitive-fluorescent} can be schematised as a low-pass filter on the activity signal.} that determines the light emission. Indeed, even if we used the fast variant (GCaMP6f), the dynamics of light emission is slower than the one of state transition. This is also the reason why we cannot estimate the slope of the Down-to-Up state transition from the profile of the function that rises from the minimum, because the fast slope is hidden in the slow variation of the transfer function. 
We assume, instead, that it is possible to recognize the trigger of the transitions 
in the signal's minima of each pixel. This hypothesis is justified by the transfer function having a characteristic rise time $\tau_{\mathrm{rise}} \sim 150 ms$ \cite{Chen2013:Ultrasensitive-fluorescent}, comparable with the theoretical minimum time interval between the passage of two waves on a pixel\footnote{This assumption, made for the wave rise time, is not valid for the decay time because the transfer function's decay time is $\tau_{\mathrm{decay}} \simeq 500ms$ \cite{Chen2013:Ultrasensitive-fluorescent}}. 
Indeed, assuming two different waves emitted in antiphase with a temporal distance of $250\ ms$ (the maximum frequency of waves in the delta band is $4\ Hz$), the same neuronal population might be involved by the passage of two wavefronts in just $125\ ms$. Taking into account this borderline case, the actual sub-band where the phenomenon is observable is $[0.5, 3.3]\ Hz$\footnote{As better explained afterward, the deep anesthesia state causes the signal of interest to belong to this frequency sub-band.}. As long as the studied signal is part of this sub-band, the technique we developed is able to detect the down- to up-state transition dynamics of a neuronal population throughout its whole duration. 
In detail, as a wave is passing on a population, the corresponding pixel's signal is already decaying: the wave passage causes its ascent and, therefore, a local minimum of the signal, that can be identified with the onset of a new transition. According to this hypothesis, it is thus possible to catch the quick upward transitions relying on optical techniques characterized by fluorescence response times much slower than the transition times themselves.

The analysis of minima can be split into two steps:
\begin{itemize}
	\item \textbf{Search of minima}. For each pixel, the signal's minima are identified, comparing the intensity value at each instant (frame) to its previous and next one. 
	This operation corresponds to evaluate the difference quotient of the data, 
	and it used as the starting point for the subsequent step of interpolation. 
	\item \textbf{Parabolic interpolation of minima}. A parabolic fit is evaluated around each minimum\footnote{For the parabolic fit, 5 data points are used. Analytically, if a fit with a second order polynomial is tried (3 parameters), 5 points are enough for the estimate of the fit parameters. We evaluated also the options of using a higher order polynomial (a quartic, 5 parameters), for which 5 points are theoretically enough, but practically 7 or 9 would be requested for a more accurate estimate of the function profile. However, under the assumption that each minimum corresponds to the passage of a wave, and that distinct waves are separated in time at least by $250\ ms$ (neglecting the bordeline case of two waves in antiphase), using 5 points (\ie inspecting $160\ ms$ around the minimum, with the sampling time of $40\ ms$) should allow to isolate each wave contribution, while with 7 or 9 points ($240\ ms$ and $320\ ms$ respectively) an overlap of more than one wave cannot be excluded.}. The three parabola's parameters are saved into a proper data structure. The time of the minimum (interpolated between original frames) is obtained from the vertex of the parabola. With this information, assumed as the timestamp of the transition, it is possible to reconstruct the activity of the passing waves, following the movement both in time and space of the minima, as we will show in \textit{Results} (Section \ref{sec:results}).
\end{itemize}

Once the collection of minima of the whole dataset is obtained, it is partitioned into slow waves. The collection of minima is given as an input to a MATLAB\footnote{MATLAB\textsuperscript\textregistered, The MathWorks, Inc., Natick, Massachusetts, United States, \url{https://www.mathworks.com/products/matlab.html} (accessed on Nov. 2019).} pipeline, already employed to analyze and elaborate data from electrodes \cite{Ruiz-Mejias2011:Slow-and-fast-rhythms, Mattia2013:MotorPlanningPremotorCortex, Ruiz-Mejias2016:Dyrk1A, Capone2017:Slow-Waves-Cortical-Slices} and recently revised and refined \cite{DeBonis2019:SWAP}.
On this, it is notable that the analysis pipeline that we devised nicely connects, after only minor adaptation efforts, with a well established analysis workflow, that was developed for addressing the study of completely different experimental data. This fact is already a result, since it represents a first successful step towards a plan of delivering a set of tools and methods, to extract quantitative information from electrophysiological signals, and to compare experiments, overcoming differences in the data taking and any systematic effects that may eventually depend on the data acquisition techniques. 

Concerning this pipeline -- that we call \textit{WaveHunt} -- its action consists in splitting the set of transition times into separate waves. In order to do this, it is necessary to set an initial maximum time interval (\emph{Time Lag}) between transitions, so that they can be labelled as part of the same wave. This value is iteratively reduced by $25\%$ until every identified wave respects the unicity principle: every pixel cannot be involved more than once by the passage of a single wave. In addition, we adopt a globality principle: we only keep waves for which at least the $75\%$ of the total pixels are recruited (\emph{global waves}). This selection is done in order to guarantee that what we call ``wave'' is actually a global collective phenomenon on the cortex\footnote{The procedure that returns the collection of waves from the set of trigger times at given positions on the cortex (channels or pixels) is a crucial issue. Criteria have to be defined ``to cut'' the time series, in order to segment the list of trigger times into waves. The unicity principle is an option, together with the globality request; of course, these assumptions select a subset of waves (global waves; predominance of regular propagation patterns, as planar and spherical wavefronts) and rule out overlapping patterns and partial waves. 
In other words, in the attempt of disentangling a complex phenomenon like the propagation of delta waves, our choices focus on extracting regular patterns, assuming they constitute a relevant part of the picture.\label{note:WaveHunt}}. 

\vspace{0.2cm}

\begin{figure*}
    \centering
    \resizebox{15.5cm}{22cm}{
    \begin{tikzpicture}[node distance = 2cm]
        \node (start) [startstop] {Import raw \textit{.tif} images as float matrices};
        \node [right = 4.0cm of start, draw = none] (enter) {};
        \node [process, below = 0.5cm of start] (init) {$\bullet$ Image cropping \\ $\bullet$ Background subtraction \\ $\bullet$ Image masking};
        \draw [arrow, solid] (start) -- (init);
        \node [process, below = 0.5cm of init] (sp_smooth) {Spatial smoothing: 2x2 block reduction};
        \draw [arrow, solid] (init) -- (sp_smooth);
        \node [process, below = 0.5cm of sp_smooth] (te_smooth) {$\bullet$ Fourier spectral analysis\\ $\bullet$ Temporal smoothing: Butterworth filter};
        \draw [arrow, solid] (sp_smooth) -- (te_smooth);
        \begin{scope} [on background layer]
            \node [draw = black, thick, fill = black!15, fit = {(start)(init)(sp_smooth)(te_smooth)}, inner sep = 10pt] (step1) (step1) {};
        \end{scope}
        \node [above, inner sep = 2pt, fill = white] at (step1.north) {\bf{IMAGE INITIALIZATION $\&$ SIGNAL SMOOTHING}};
        \node [process, text width = 3cm, below = 5.7cm of init] (minima) {$\bullet$ Search of minima \\ $\bullet$ Parabolic interpolation of minima};
        \draw [arrow, solid] (te_smooth) -- (minima);
        \node [process, below = 0.5cm of minima] (wavehunt) {WaveHunt};
        \draw [arrow, solid] (minima) -- (wavehunt);
        \begin{scope} [on background layer]
            \node [draw = black, thick, fill = black!15, fit = {(minima)(wavehunt)}, inner sep = 10pt] () (step2) {};
        \end{scope}
        \node [above, inner sep = 2pt, fill = white] at (step2.north) {\bf{FIND UPWARD TRANS. TIMES $\&$ WAVES CONSTRUCTION}};
        \node [process, text width = 3cm, below = 3cm of minima, xshift = -5cm] (excitability) {Excitability};
        \draw [arrow, solid] (minima) -- node[near start, above, fill = white,] {} + (-5, 0) -- (excitability);
        \node [process, below = 3cm of minima] (origins) {Origin of waves};
        \draw [arrow, solid] (wavehunt) -- (origins);
        \node [process, below = 3cm of minima, xshift = 5cm] (speed) {Speed of waves};
        \draw [arrow, solid] (wavehunt) -- node[near start, above, fill = white,] {} + (+5, 0) -- (speed);
        \begin{scope} [on background layer]
            \node [draw = black, thick, fill = black!15, fit = {(excitability)(origins)(speed)}, inner sep = 10pt] () (step3) {};
        \end{scope}
        \node [above, inner sep = 2pt, fill = white] at (step3.north) {\bf{NEURONAL ACTIVITY MEASURES}};
    \end{tikzpicture}}
    \caption{\small Flowchart of the analysis pipeline.} \label{fig:Flow_Chart_Analisi}
\end{figure*}
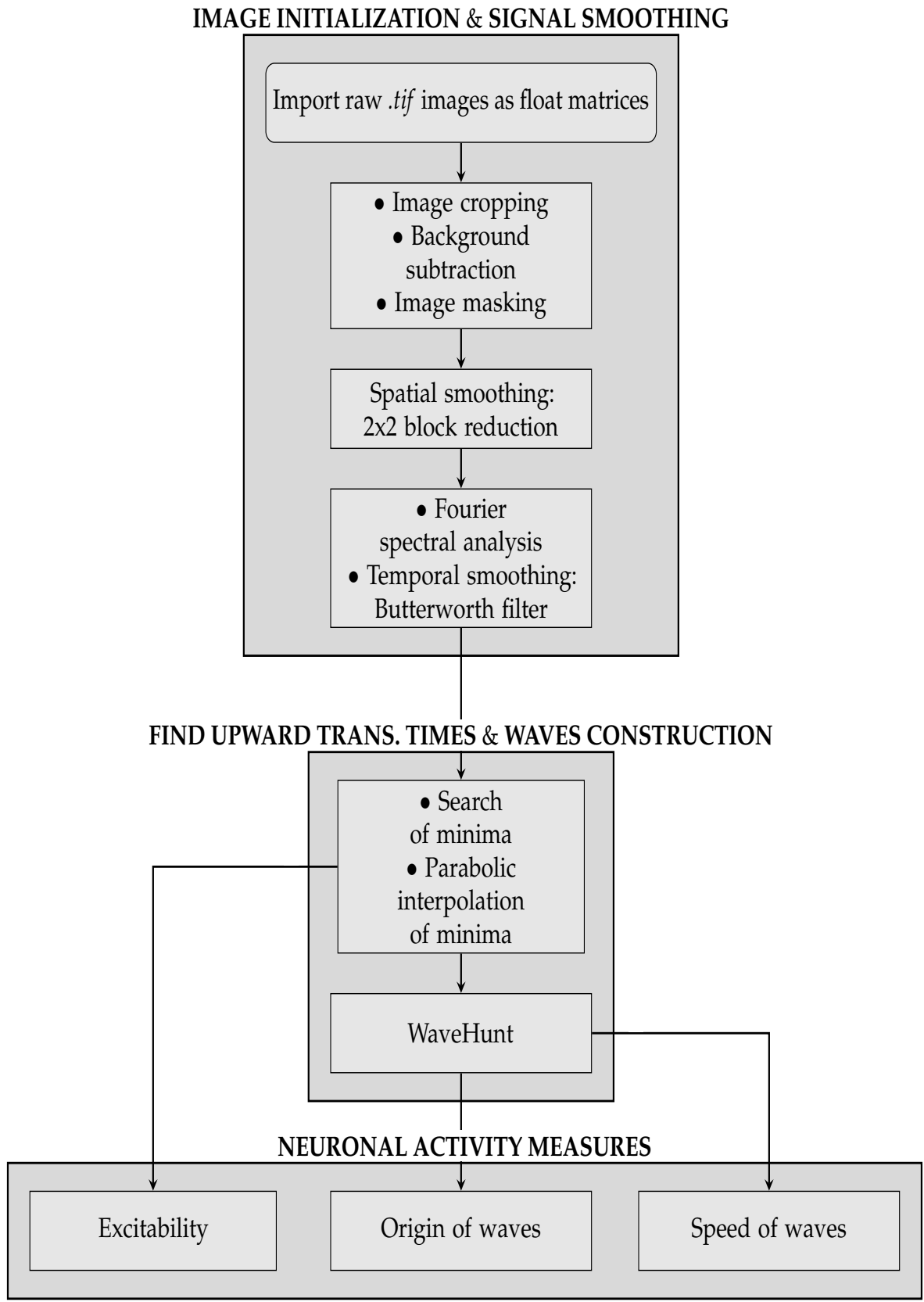

With the collection of waves, it is now possible to get to the third part of the analysis pipeline, where quantitative information on the slow wave activity is obtained. In this phase, three different types of measures are taken: 
\begin{itemize}
    \item \textbf{Excitability of the neuronal populations}. For each minimum, the quadratic coefficient of the corresponding parabolic interpolation is taken. 
    It is proportional to the concavity of the parabola, therefore it contains information on the excitability that the respective neuronal population (\ie pixel) exhibits as a consequence to the wave that is passing\footnote{In detail, the quadratic coefficient is proportional to the second derivative of the function, and it is a measure of the speed of variation of the first derivative; the first derivative indicates a variation of the function; the function is a measure of the luminosity, and it is related to the state of the tissue (low/high emitting because of the variation in the calcium concentration). The first derivative can identify the minima of the function, and thus the timestamps at which a most significant variation (a change in the state, a Down-to-Up transition) is occurring; the fastest is such variation (monitored by the values of the second derivative), the fastest is the state transition, and thus, in our description, the most reactive (\ie. excitable) is the tissue underneath.}; 
    the unit of the excitability is $\left[t^{-2}\cdot\frac{\mathcal{F}}{\mathcal{F}_{max}}\right]$. 
    In order to study the distribution of excitability, values of excitability are collected within a histogram reporting statistics of the entire sample of minima. 
    In addition to this, an \emph{Excitability Map}, representing the average excitability for each pixel, is produced\footnote{We point out that the information on the excitability can be obtained without splitting the collection of minima into global waves, \ie before the WaveHunt step of the pipeline, that is necessary only for origin points and velocity. This is clearly shown in the flowchart at the end of this section (Figure \ref{fig:Flow_Chart_Analisi}), in which the logic of the whole pipeline is represented in a schematic form.}.
	\item \textbf{Origin Points of the waves}. Once that the collection of waves is obtained, each wave in the collection is examined separately. For each wave, there is a sequence of pixel activation; the first $N$ pixel in the sequence are identified as the origin points\footnote{We set $N=30$ in this work; this value, given the number of non-NAN pixels in the images and the request of globality set at $75\%$, corresponds to having the subset of origin points with less than $3\%$ of the pixels involved in each wave.}. Some pixels more often appear in this ranking, and are the ones that identify the dominant spatial origin of the waves. 
	\item  \textbf{Wave velocity}. In order to obtain the average speed of each wave, the wave velocity is calculated on different points. If the passage time function\footnote{The passage time function indicates, for each position $(x, y)$, the time at which that position has been reached by the wave during its propagation.} of the wave $T(x, y)$ is known, the speed of a wave on a point $(x, y)$ can be defined as the inverse of the module of the function's gradient, 
	\begin{equation}\label{eq:speed_gradient}
		v(x,y) = \frac{1}{|{\nabla} T(x,y)|}.
	\end{equation}
	
    Computing the gradient and taking its module, we obtain:
	\begin{equation} \label{eq:speed_component}
		v(x,y) = \frac{1}{\sqrt{(\frac{\partial T(x,y)}{\partial x})^2 + (\frac{\partial T(x,y)}{\partial y})^2}}.
	\end{equation}
	It should be noted that the function $T(x, y)$ represents the wave passage time (that we identify with the time of the minima) in the spatial continuum, bur we have this information only for the discrete points corresponding to the pixels. For this reason, the partial derivatives that appear in \eqref{eq:speed_component} have been calculated as finite differences, with the distance between two pixels denoted as $d$:
		
	\begingroup\makeatletter\def\f@size{8.7}\check@mathfonts
		\def\maketag@@@#1{\hbox{\m@th\normalsize\normalfont#1}}
		\begin{equation} \label{eq:speed_component_discrete-points}
	        \left\{\begin{array}{ll}
		    \dfrac{\partial T(x,y)}{\partial x} \simeq \dfrac{T(x + d,y) - T(x - d,y)}{2d}\\ \\
		    \dfrac{\partial T(x,y)}{\partial y} \simeq \dfrac{T(x ,y + d) - T(x ,y - d)}{2d}.
		    \end{array}
		    \right.
		\end{equation}
	\endgroup
		
	In other words, for any pixel identified by the pair of coordinates $(x, y)$, the velocity of the wave on that pixel is computed using the information collected on the four adjacent pixels. Therefore, given the above calculation rule, for each wave only pixels with four valid neighbours are taken into account: these are pixels for which it is possible to evaluate both the $\frac{\partial T(x, y)}{\partial x}$ and the $\frac{\partial T(x, y)}{\partial y}$. By averaging the values obtained over pixels that meet this condition, we obtain the average speed of a wave; this procedure is repeated for each wave, in order to create a histogram of average speeds.
	\end{itemize}
	
Through the measures above listed, we have been able to quantitatively compare the data coming from the two different mice, and from the simulation obtained with the Toy Model.
A flowchart of the whole analysis pipeline is presented in Figure \ref{fig:Flow_Chart_Analisi}.

\let\clearpage\relax

\subsection{Toy Model}
\label{subsec:ToyModel}

The theoretical model underlying the realization of the simulator is based on simple conceptual assumptions. The system is schematized as a two-dimensional grid of \emph{Pixels}\footnote{Python objects that define the elements of the simulation are indicated in \emph{italic}.}; width and height of the grid can be easily set to reproduce the shape and size of the experimental images. Each \emph{Pixel}, being ideally the representation of a macroscopic portion of the cerebral cortex, contains a number of \emph{Neurons} extracted from a normal distribution $\mathcal{N} \left(\mu = 10, \sigma = 2 \right)$\footnote{The obtained number of \emph{Neurons} per \emph{Pixel} is unrealistic and underestimates the density of cells in the assembly, but it is requested for facing the limited processing resources of the simulation; in addition, it can allow a faster computation. Nevertheless, the impact of this choice is not relevant for the results of the Toy Model, since the \emph{Pixel}'s signal is normalized.}.
\emph{Neurons} are the actual sources of the signal, as Poissonian emitters. 
In the simplest theoretical modelling, only two options are admitted for the expected value of the Poisson distribution that emulates the spiking activity of the \emph{Neurons}: one for the Down state (expected value set at $\mu_{\mathrm{Down}}$), and the other for the Up state (expected value set at $\mu_{\mathrm{Up}}$). The simplest model has \emph{Neurons} as independent, identical units, each holding an internal binary state that identifies the \emph{Neuron} as active (Up) or idle (Down). 
The internal state of each \emph{Neuron} depends only on the state of the \emph{Pixel} to which the \emph{Neuron} is assigned. It is straightforward that the \emph{Pixel} can be in two states: active (if invested by a \emph{Wave}), or idle (if no \emph{Wave} has passed by the \emph{Pixel}'s coordinates for a given time interval). If the \emph{Pixel} is active, its \emph{Neurons} are Up; if the \emph{Pixel} is in idle, its \emph{Neurons} are Down. Consequently, the Poissionian emitters of that \emph{Pixel} will fire according to the corresponding expected value ($\mu_{\mathrm{Up}}$ and $\mu_{\mathrm{Down}}$, respectively). 

Experimental evidences \cite{DeBonis2019:SWAP}
indicate that the ratio of the active state firing rate $\mu_{\mathrm{Up}}$ to the idle state firing rate $\mu_{\mathrm{Down}}$ can be estimated within the interval: $\frac{\mu_{\mathrm{Up}}}{\mu_{\mathrm{Down}}}\in \left[4, 8 \right]$. Considering these results, it is therefore reasonable to set this ratio at $5$ for the Toy Model\footnote{This assumption is simplistic but rooted in experimental observations: in \cite{DeBonis2019:SWAP}, we compute the distribution of $log(MUA)$, the natural logarithm of the multi-unit activity, for $11$ murine subjects, $32$ channels per subject and, despite the large variability of the sample, we observe that the activity in the Up state is between $\mathrm{e}^1$ and $\mathrm{e}^2$ times larger than in the Down state ($\mathrm{e}=2,718...$ is Euler's number). }.

Each \emph{Neuron} has an intrinsic time of permanence in the active state, $\tau_{\mathrm{Up}}$, after which it returns to the idle state. To fix this parameter -- univocal, for simplicity, for the entire population of \emph{Neurons} -- we looked at experimental observations that place the duration of the Up state in a range $ \tau_{\mathrm{Up}} \in \left[150, 200 \right]\ ms$ \cite{DeBonis2019:SWAP}; 
in the Toy Model, we set $\tau_{\mathrm{up}} = 200\ ms $. A \emph {Pixel} is declared inactive when all its \emph{Neurons} are turned in idle state.

In order to model the presence of the fluorescent protein, from which the luminosity signal is obtained, the spiking activity of each \emph{Pixel} undergoes a convolution with a response function. The analytical form of the response is derived from the inspection of results reported in \cite{Chen2013:Ultrasensitive-fluorescent}, aiming at reproducing the trend of the fast variant (GCaMP6f); the function used in the Toy Model is a Lognormal, defined as:

\begingroup\makeatletter\def\f@size{8.5}\check@mathfonts
    \def\maketag@@@#1{\hbox{\m@th\normalsize\normalfont#1}}
    \begin{equation}
        \mathrm{Lognormal} \left(x; \mu, \sigma \right) = \frac{1}{x} \frac{1}{\sqrt{2 \pi} \sigma} \exp \left( - \frac{\left( \ln x - \mu \right)^2}{2 \sigma^2} \right).
        \label{eq:Lognormal-def}
    \end{equation}
\endgroup

This function depends on the set of parameters $\mathcal{P} = \left\{ \mu, \sigma \right\}$ and its shape varies with them.
A first and significant constraint that the response function must respect is the agreement between 
the experimentally observed rise time $\tau_{\mathrm{rise}}$ and the modelled one. Looking at results shown in reference \cite{Chen2013:Ultrasensitive-fluorescent}, 
it is possible to identify a range of validity, $\tau_{\mathrm{rise}} \in \left[150, 200 \right] \ ms$, while the modelled value can be taken as the mode $\mathcal{M}$ of the function, resulting in the following condition: 

\begin{equation}
    \mathcal{M} \Big[ \mathrm{Lognormal} \Big] = \exp \left( \mu - \sigma^2 \right) = \tau_{\mathrm{rise}}.
    \label{eq:Mode-Lognormal}
\end{equation} 

In order to be able to univocally identify the parameters $\mu$ and $\sigma$, it is necessary to impose an additional constraint, for example a moment of the distribution. However, given the difficulty of extracting such a piece of information from the visual inspection of the experimental response, it was preferred to explore a set of reasonable values compatible with the constrain expressed in \eqref{eq:Mode-Lognormal} and capable of reproducing as closely as possible the immediately visible characteristics of the function.

\begin{figure}[h]
	\centering
    \begin{overpic}[scale = 0.47]{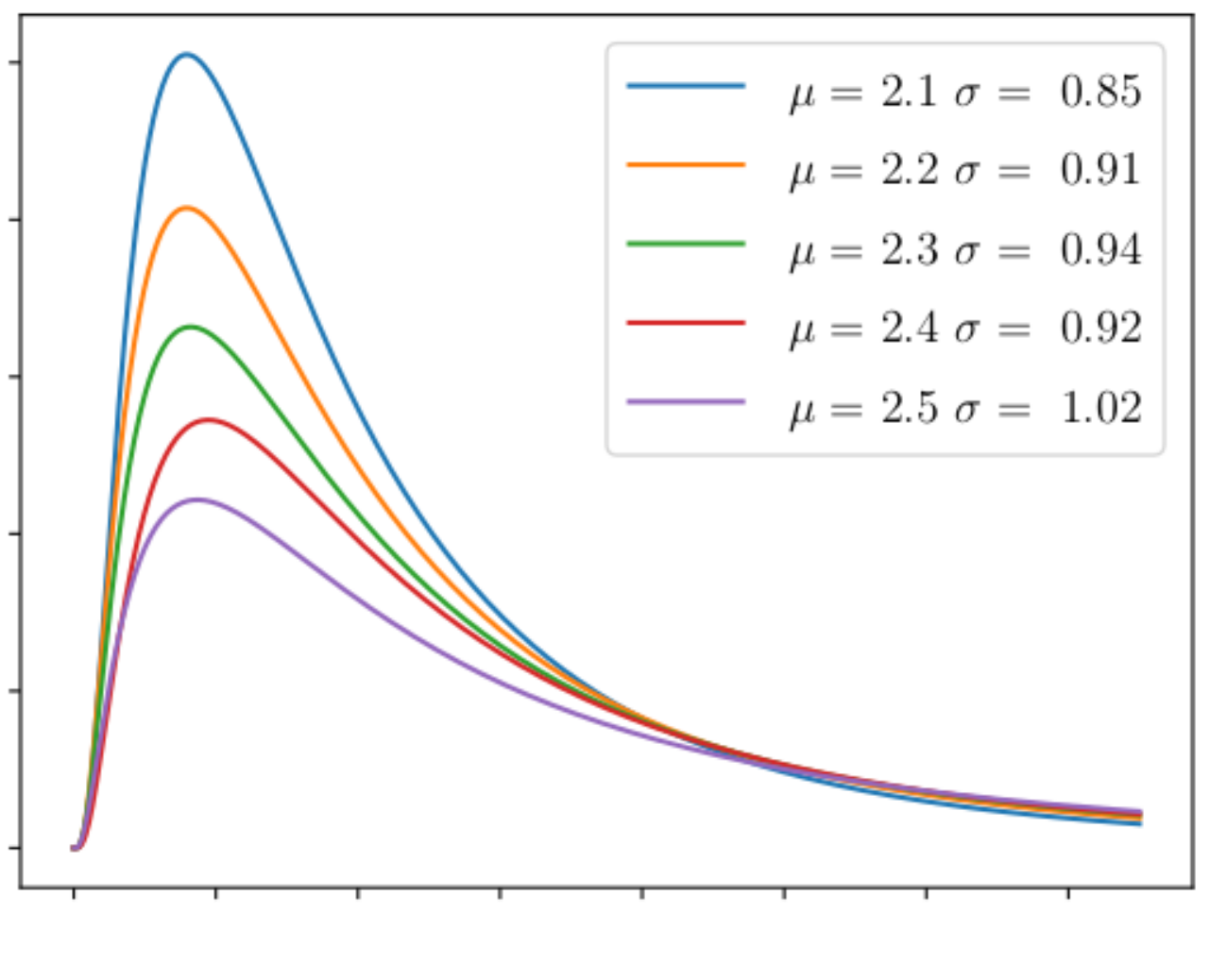}
        \footnotesize \put (4, 3) {$0.0$}
        \footnotesize \put (16, 3) {$0.2$}
        \footnotesize \put (28, 3) {$0.4$}
        \footnotesize \put (40, 3) {$0.6$}
        \footnotesize \put (51, 3) {$0.8$}
        \footnotesize \put (63, 3) {$1.0$}
        \footnotesize \put (74, 3) {$1.2$}
        \footnotesize \put (86, 3) {$1.4$}
        \footnotesize \put (-9, 10) {$0.00$}
        \footnotesize \put (-9, 23) {$0.05$}
        \footnotesize \put (-9, 36) {$0.10$}
        \footnotesize \put (-9, 49) {$0.15$}
        \footnotesize \put (-9, 62) {$0.20$}
        \footnotesize \put (-9, 75) {$0.25$}
        \put (45, -2) {\small Time $\left[ s  \right]$}
        \put (-18, 20) {\small \rotatebox{90}{Response function $\frac{\mathcal{F}}{\mathcal{F}_0}$}}
    \end{overpic}
    \caption{\small Lognormal functions for different set of parameters $ \mu $ and $ \sigma $. The pair $\left\{\mu = 2.2, \sigma = 0.91 \right\}$ (in orange) is selected for the emulation of the response of GCaMP6f.}
    \label{fig:ToyModel_LogNorm_Trends}
\end{figure}
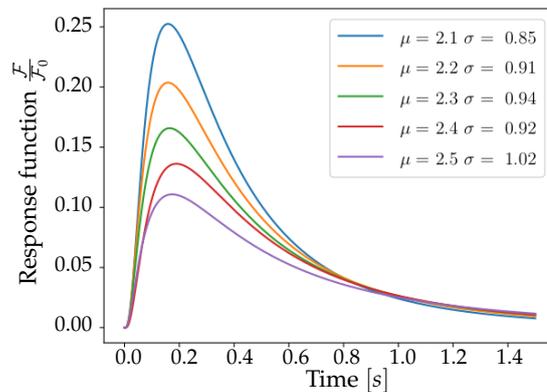

In Figure \ref{fig:ToyModel_LogNorm_Trends}, a collection of Lognormal trends is plotted for different $\left\{ \mu, \sigma \right\}$ values that satisfy \eqref{eq:Mode-Lognormal}. Trends are normalized to $\mathcal{F}_0$,the median of the distribution -- in order to facilitate the comparison between reports in \cite{Chen2013:Ultrasensitive-fluorescent} 
and Figure \ref{fig:ToyModel_LogNorm_Trends}. This comparison shows important characteristics of the pair $\mathcal{P}^\star \equiv \left\{ \mu = 2.2, \sigma = 0.91 \right\}$: the corresponding function, when decaying, halves its maximum value after a time interval of $t_{\mathrm{half}} = 0.5\ s$, and reaches zero after about one second, mimicking the properties of the curve reported by \cite{Chen2013:Ultrasensitive-fluorescent} for the 6f variant of GCaMP. 
From these considerations, it emerges that the pair of parameters $\mathcal{P}^\star$ is a valid choice for the description of the experimental response function, being able to reproduce the immediately observable features of amplitude, asymmetry and tail weight. This set $\mathcal{P}^\star$ has thus been used for the entire population of \emph{Neurons}.

Once the simulation of Poissonian emitters is completed, each \emph {Pixel} collects the sum of the signals produced by \emph {Neurons} belonging to it. This population signal is then convolved with the above discussed Lognormal, to generate the fluorescence response, resulting in the brightness signal of the whole \emph{Pixel}. 


Since the \emph{Neurons} described in the Toy Model are intrinsically independent objects, no form of wave propagation or interaction would be possible. The dynamics is therefore implemented separately according to two different conceptual schemes.
A first possibility consists of defining a certain number of waves during the simulation initialization. A wave is nothing more than an object defined on the two-dimensional grid with its own parameters of shape, position, velocity and acceleration. For each instant of time, the simulation evaluates the correct wave dynamics (processed according to the parameters previously specified) and checks the spatial overlap of waves and \emph{Pixels}, assigning the activation status to each \emph{Pixel}. Each wave is equipped with further parameters in order to make its dynamics possibilities more flexible and broad; these parameters include an initial position of birth, a life time, a condition of reflection or absorption at the edges of the grid.
A second possibility, developed for a more accurate comparison with the experimental data, consists of declaring a sequence of activation times for each \emph{Pixel}. The simulation proceeds checking step by step which \emph{Pixel} must be activated to respect the time sequences given in input.

\begin{figure*}
    \centering
    \resizebox{15.5cm}{21cm}{
    \begin{tikzpicture}[node distance = 2cm]
        \node (start) [startstop] {Start};
        \node [right = 4.0cm of start, draw = none] (enter) {};
        \draw [arrow] (enter) -- node[anchor=south] {Input UpFile Trans} (start);
        \node [process, below = 1.0cm of start] (init) {Initialize Simulation:\\\vspace{0.1cm} $\bullet$ \emph{initFromFile}\\  or \\ $\bullet$ \emph{initWithWaves}};
        \draw [arrow, solid] (start) -- (init);
        \node [process, text width = 3cm, below = 3cm of init] (upN_generate) {Generate Poisson\\ Signal};
        \draw [arrow, solid] (init) -- (upN_generate);
        \node [process, below = 1cm of upN_generate] (upN_convolve) {Store Neuron Signal};
        \draw [arrow, solid] (upN_generate) -- (upN_convolve);
        \begin{scope} [on background layer]
            \node [draw = black, thick, fill = black!15, fit = {(upN_generate)(upN_convolve)}, inner sep = 10pt] (updateNeron) (updateNeuron) {};
        \end{scope}
        \node [above, inner sep = 2pt, fill = white] at (updateNeuron.north) {\bf{UPDATE NEURON} $j$};
        \node [decision, aspect = 2, below = 1cm of upN_convolve] (exit_upN) {Are neurons finished?};
        \draw [arrow, solid] (updateNeuron) -- (exit_upN);
        \draw [arrow, solid] (exit_upN.east) -- node[near start, above, xshift = 0.40cm, yshift = 0.02cm,  fill = white] {No} + (1.1, 0) |- node[near start, right] {$j = j + 1$} (updateNeuron.east);
        \begin{scope} [on background layer]
            \node [draw = black, thick, fit = {(updateNeuron) (exit_upN)}, inner sep = 20pt] (updatePixel) {};
        \end{scope}
        \node [above, inner sep = 4pt, fill = white] at (updatePixel.north) {\bf{UPDATE PIXEL} $i$};
        \node [decision, aspect = 2, below = 0.5cm of updatePixel] (exit_upP) {Are pixels finished?};
        \draw [arrow, solid] (exit_upN.south) -- node[near start, right] {Yes} (exit_upP);
        \draw [arrow, solid] (exit_upP.west) -- node[near start, above, fill = white, xshift = -0.4cm, yshift = 0.03cm] {No} + (-1.4, 0) |- node[near start, left, xshift = 0.5cm, yshift = 3.3cm] {$i = i + 1$} (updatePixel.west);
        \node [draw = black, dashed, fit = {(updatePixel) (exit_upP)}, inner sep = 20pt, minimum width = 10.5cm] (stepSimulation) {};
        \node [above, inner sep = 4 pt, fill = white] at (stepSimulation.north) {\bf{UPDATE TIME} $t$};
        \node [decision, aspect = 2, below = 1 cm of exit_upP] (time_over) {Is $t \ge t_{end}$?};
        \draw [arrow, solid] (time_over.west) -- node[near start, above, fill = white, yshift = 0.03cm, xshift = -0.7 cm] {No} + (-4, 0) |- node[near start, left] {$t = t + dt$} (stepSimulation.west);
        \draw [arrow, solid] (exit_upP.south) -- node[near start, right] {Yes} (time_over.north);
        \node [io, right = of time_over, text width = 3cm] (save) {For all Pixels: Convolve the Signal and Save the Results};
        \node [startstop, below = 1cm of save] (stop) {Stop};
        \draw [arrow, solid] (time_over) -- node[near start, above] {Yes} (save);
        \draw [arrow, solid] (save) -- (stop);
    \end{tikzpicture}}
    \caption{\small Flowchart of the simulation.} \label{fig:Flow_Chart_Simulazione}
\end{figure*}
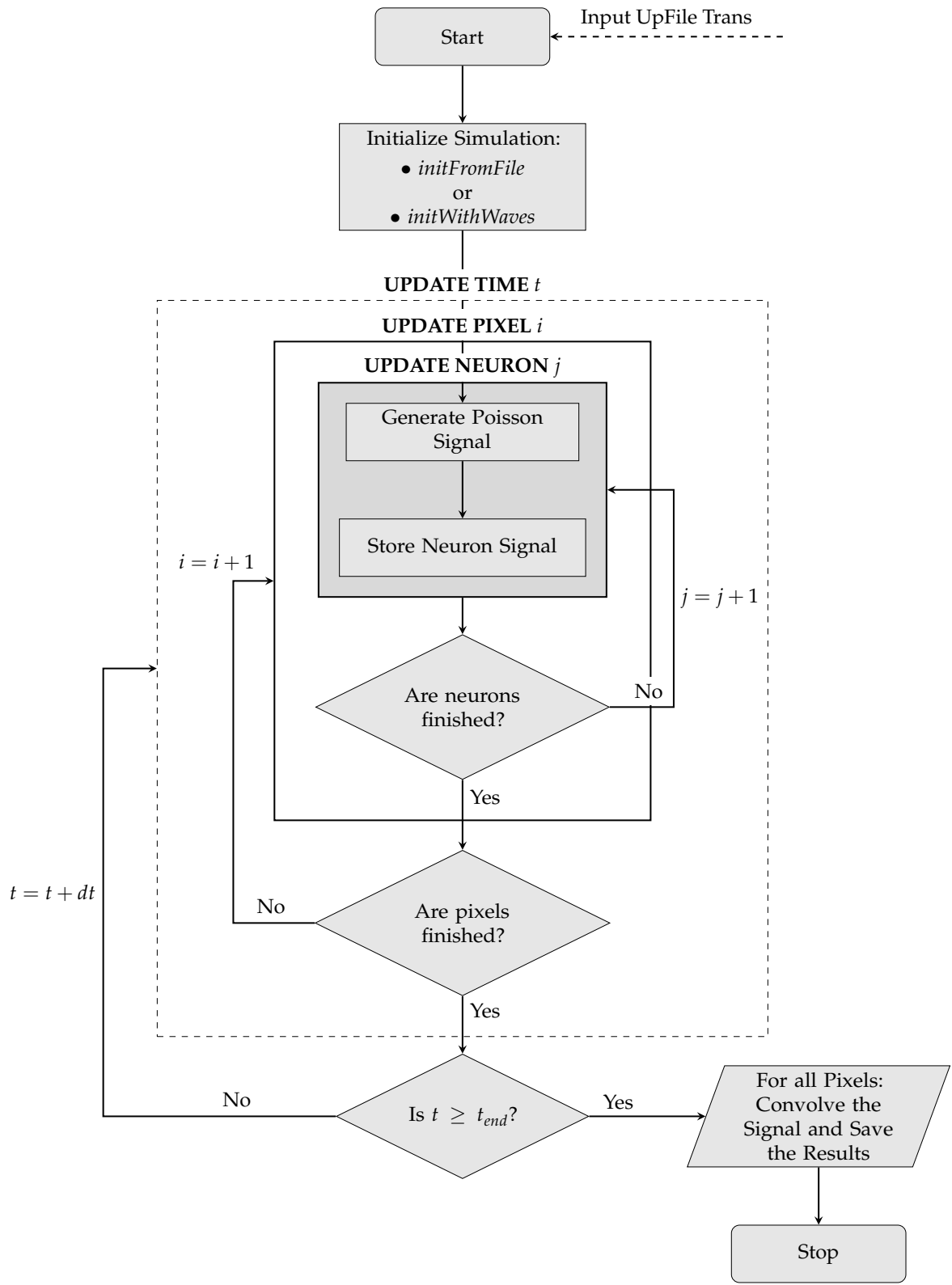

On the next page, the logic implemented in the simulation -- starting from the initialization phase up to the production of the final signal -- is schematized in the form of a flowchart (Figure \ref{fig:Flow_Chart_Simulazione}).

\let\clearpage\relax
The Toy Model is not intended to provide a complete picture of the light emission phenomenon, given its complexity and the fact that the fluorescence signal in the wide-field Calcium imaging is the result of the interplay of several factors, not easily disentangled. The signal could indeed generate from neurites and somas of pyramidal neurons from both layer 2/3 and layer 5, and it was indeed reported that the neuronal firing is one of the components of the observed emission, but a large fraction of the fluorescence signal would originate in the neuropile, reporting on the activation of dendritic and axonal Calcium channels \citep{Yoshida2018:Ca2Efflux}.  
%
%
We decided to focus on the neuronal firing component only, for three reasons: (i) it is the most straightforward approach to follow, enabled by a simplified neuron model as a point-like Poissonian emitter; (ii) in an oversimplified frame in which only one element is included, its contribution can be more clearly evaluated and emulated; in this case, since we observe a good agreement between simulated and experimental data, we can conclude that the impact of neuronal firing on the observed Ca signal is not negligible; (iii) rather than provide a valid description of the light emission phenomenon or enlarge our understanding of the cortical dynamics, our intent with the Toy Model is, less ambitiously, to offer an additional and complementary tool to calibrate and validate the analysis pipeline. The Toy Model produces an output that can emulate the data extracted from the set of images, and for the purpose of testing the stability and the validity of the analysis procedure, we care more about the format than the accurate simulation of the slow wave dynamics. The goal is to feed the pipeline with two separate and independent inputs, aiming at attesting its flexibility and its potential to operate on different datasets without the need of carrying out a custom tuning of the procedure; the results we obtain on this are encouraging, as discussed in the following Section.

\section{Results}
\label{sec:results}

In this Section, we describe the application of the new analysis pipeline (described in Section \ref{subsec:DataAnalysisPipeline}) on wide-field images (presented in Section \ref{subsec:Methods-DAQ}). 
In parallel with the data analysis, we developed a Toy Model to simulate the neuronal activity extracted from the experimental images (Section \ref{subsec:ToyModel}). 
The hypothesis that we want to validate here is that the combined and integrated activity of many Poissonian variables, convolved with an appropriate transfer function and subjected to a modulation of the emission frequency (occurring at proper times, extracted from experimental data), can reproduce signals whose shape is similar to that observed in experimental recordings.
With this aim, the synthesized signals are processed through the same analysis pipeline developed for the treatment of the experimental data. A general consistency between the results obtained from the experimental samples and from the simulated ones can be interpreted as a validation of the analysis procedure. 

\subsection{Analysis of the Experimental Data}
Raw images are imported as matrices of floating-point numbers, cropped and masked, then the background is subtracted and the spatial smoothing is performed (macro-pixels); these steps are illustrated in Figure \ref{fig:Inizializzazione_Immagini}.
Moreover, in order to have the signal in normalized units, for each macro-pixel the values are divided by the absolute maximum value in the pixel ($\frac{\mathcal{F}}{\mathcal{F}_{max}}$).
The normalized signal is then analyzed in the frequency domain; the average spectrum of each set of $1000$ frames ($40\ s$) is computed as the mean of all macro-pixels spectra; a sample is shown in Figure \ref{fig:spettro_170110}.
The frequency range of interest is then selected, and the signal is cleaned up via a $6^{th}$ order band-pass Butterworth filter (time smoothing). A visual inspection of the spectra for both mice suggests to adopt a low-cut frequency $\nu_{\mathrm{low}} = 0.5\ Hz$ and a high-cut $\nu_{high} = 3\ Hz$; outside this passband, in the range
$ \nu_{\mathrm{high}} < \nu < \nu_{\mathrm{Nyq}} = 12.5\ Hz$\footnote {For the Nyquist theorem, with a sampling frequency of $\nu_{sampling} = 25\ Hz$, we are able to reconstruct frequency components up to $\nu_{Nyq} = 12.5\ Hz$.}, the spectral contribution appears small, thus associable with noise.
%
The selected band falls in the \textit{delta} band ($\left[0.5, 4 \right]\ Hz$), but $\nu_{\mathrm{high}} < 4\ Hz$, \ie the observed spectral content is shifted at lower frequencies; we believe that this due to the doses of administered anaesthetic, that induce a deep state of anaesthesia in the mice. Anyhow, $\nu_{\mathrm{high}} = 3\ Hz$ implies that the activity of interest lies in the sub-band $[0.5, 3.3]\ Hz$, in which the assumptions made in Section \ref{subsec:DataAnalysisPipeline} on the detectability of the Down- to Up-state wavefronts are valid. In Figure \ref{fig:Confronto_segnale1}, a comparison between raw signal and clean (\ie filtered) signal is presented for a given pixel.

\begin{figure}[t!]
    \centering
    \captionsetup[subfloat]{oneside,margin={-0.05cm,0cm}}
                 

    \subfloat[\emph{\emph{Raw Image}}]{
        \label{fig:Inizializzazione_Raw}
        \includegraphics[width=0.205\textwidth]{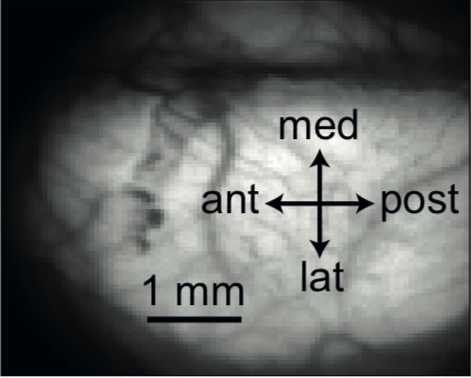}} \qquad
    \subfloat[\emph{Cutting and masking}]{
        \label{fig:Inizializzazione_Taglio-Maschera}
        \includegraphics[width=0.205\textwidth]{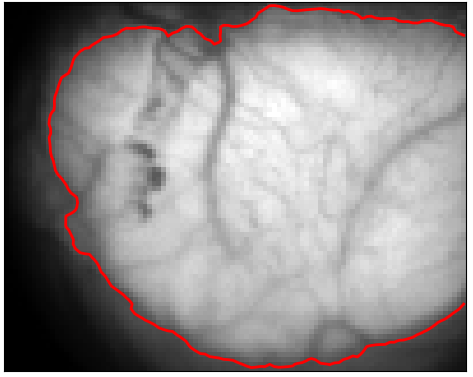}} \qquad
    \subfloat[\emph{Without background}]{\hspace{0.1cm}
        \label{fig:Inizializzazione_Background}
        \includegraphics[width=0.205\textwidth]{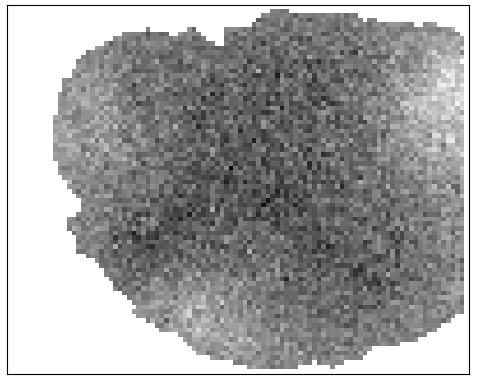}} \qquad
    \subfloat[\emph{Spatial Smoothing}]{
        \label{fig:Inizializzazione_Smoothing}	
        \includegraphics[width=0.21\textwidth]{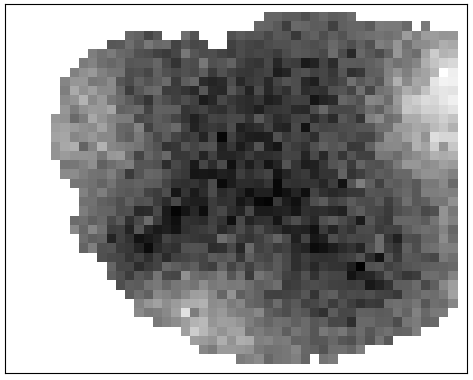}}
	\caption{\small Sequence showing the different steps of the signal cleaning process. (a) Example of a $100 \times 100$ pixels \emph{raw} image. (b) Image cropping to exclude the non-informative area: only the left hemisphere has been selected (image size: $100 \times 80$ pixels) and a mask has been created (specifically, $5564$ out of $8000 $ original pixels are within the contour, to be considered as signal sources). (c) Background, computed as the average signal, is subtracted from masked images. (d)  Spatial smoothing: blocks of $2 \times 2$ pixels constitute a macro-pixel; the number of informative channels is reduced to a factor $1/4$ of the previous amount (in this specific case, $1391$ macro-pixels are obtained).}
	\label{fig:Inizializzazione_Immagini}
\end{figure}
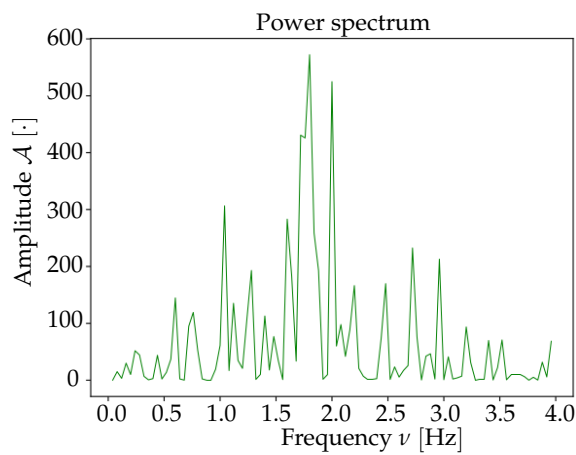
\begin{figure}[h!]
    \centering
    \begin{overpic}[width=0.42\linewidth]{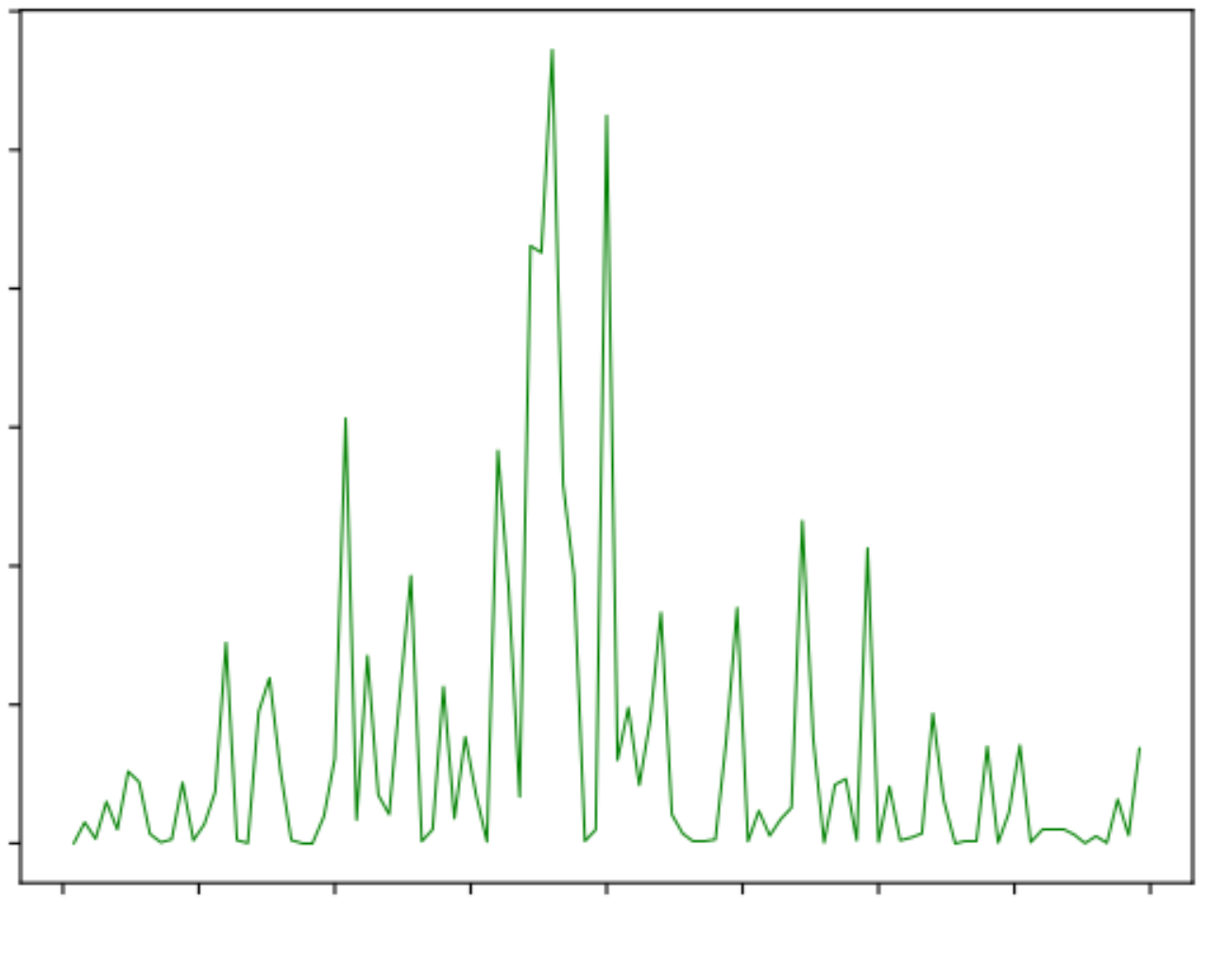}
        \small \put (3, 3) {$0.0$}
        \small \put (14, 3) {$0.5$}
        \small \put (25, 3) {$1.0$}
        \small \put (36, 3) {$1.5$}
        \small \put (48, 3) {$2.0$}
        \small \put (59, 3) {$2.5$}
        \small \put (70, 3) {$3.0$}
        \small \put (81, 3) {$3.5$} 
        \small \put (93, 3) {$4.0$}
        \small \put (40, -2) {\small Frequency $\nu$ $\left[\mathrm{Hz} \right]$}
        \small \put (-3, 10) {$0$}
        \small \put (-7, 22) {$100$}
        \small \put (-7, 33) {$200$}
        \small \put (-7, 44) {$300$}
        \small \put (-7, 56) {$400$}
        \small \put (-7, 68) {$500$}
        \small \put (-7, 79) {$600$}
        \small \put (-14, 30) {\rotatebox{90}{\small Amplitude $\mathcal{A}$ $\left[ \cdot \right]$}}
        \put (35, 82) {Power spectrum} 
    \end{overpic} \\ \vspace{0.5cm}
    \caption{\small Frequency spectrum, averaged across the pixels, for a set of $1000$ frames, corresponding to $40\ s$ of recordings (mouse referred to as \emph{Keta1}).}
    \label{fig:spettro_170110}
\end{figure}
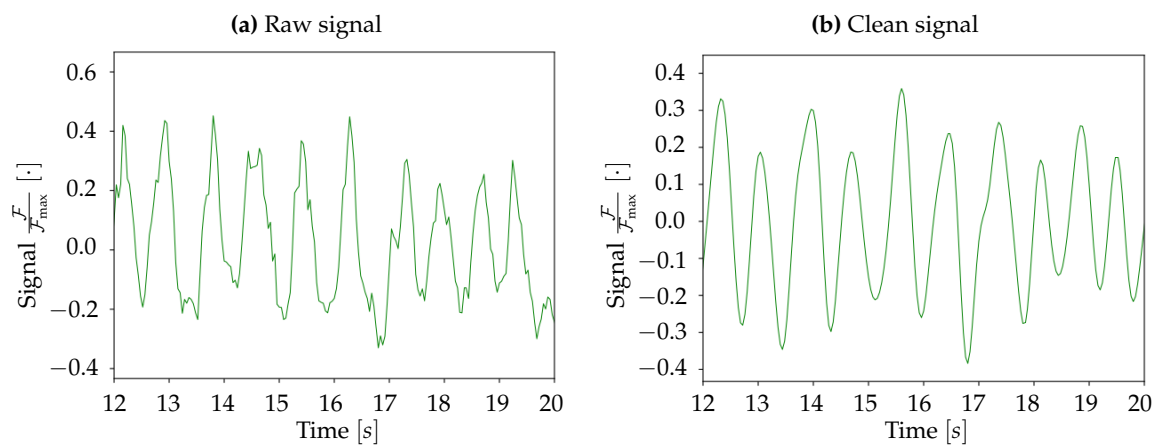
\begin{figure}[h!]
    \centering
    \captionsetup[subfloat]{position=top}
    \subfloat[Raw signal]{
        \hspace{0.5cm}
        \label{fig:Confronto_Segnale_Raw}
        \begin{overpic}[width=.38\textwidth]{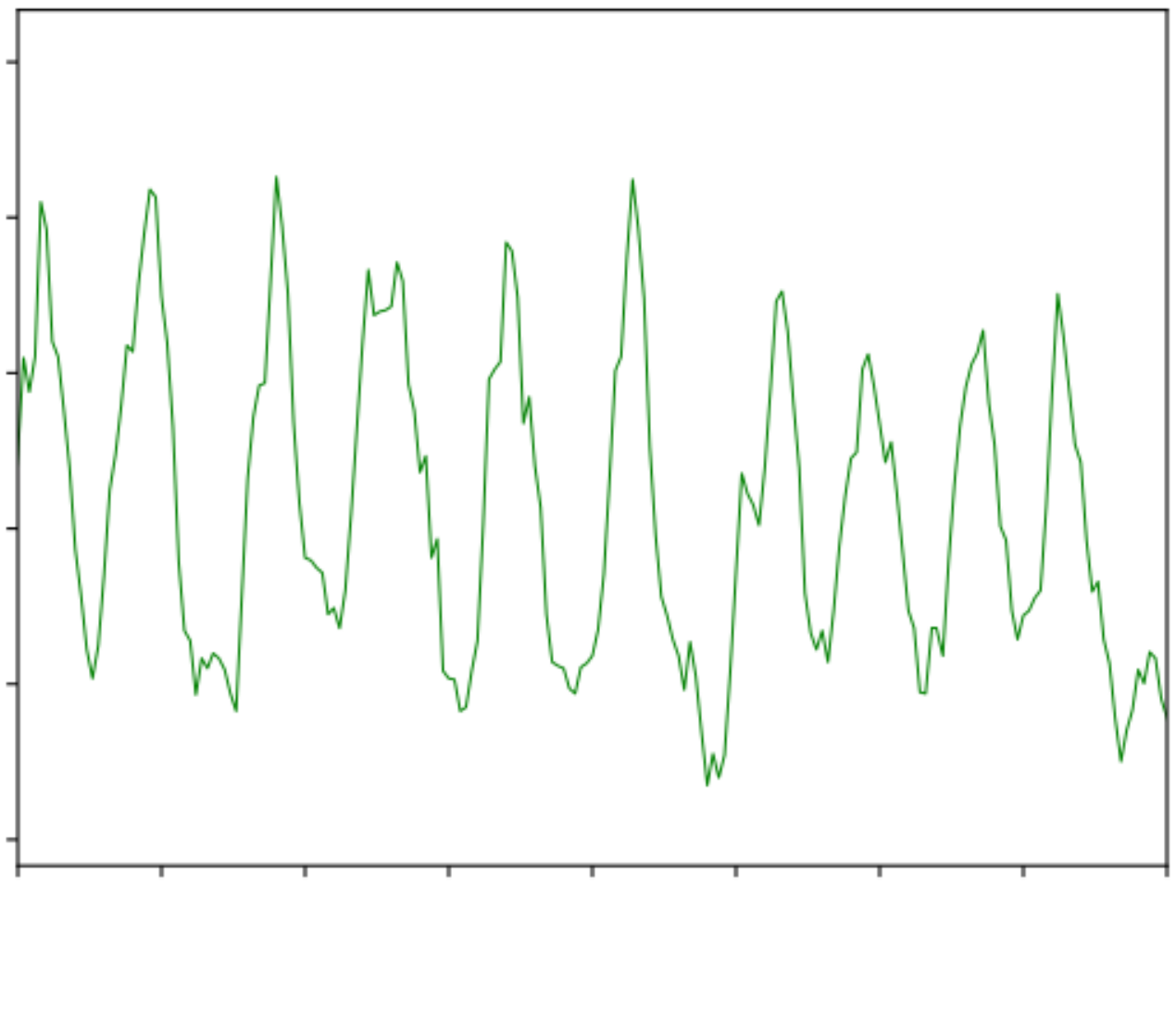}
            \put (-1, 7) {\small $12$}
            \put (11, 7) {\small $13$}
            \put (23, 7) {\small $14$}
            \put (35, 7) {\small $15$}
            \put (47, 7) {\small $16$}
            \put (59, 7) {\small $17$}
            \put (71, 7) {\small $18$}
            \put (84, 7) {\small $19$}
            \put (96, 7) {\small $20$}
            \put (42, 1) {\small Time $\left[ s \right]$}
            \put (-13, 15) {\small $-0.4$}
            \put (-13, 28) {\small $-0.2$}
            \put (-9, 42) {\small $0.0$}
            \put (-9, 55) {\small $0.2$}
            \put (-9, 68) {\small $0.4$}
            \put (-9, 81) {\small $0.6$}
            \put (-21, 30) {\rotatebox{90}{\small Signal $\frac{\mathcal{F}}{\mathcal{F}_{\mathrm{max}}}$ $ \left[ \cdot \right]$}}
        \end{overpic}} \qquad \quad \
    \subfloat[Clean signal]{
        \hspace{0.5cm}
        \label{fig:Confronto_Segnale_Pulito}
        \begin{overpic}[width=.38\textwidth]{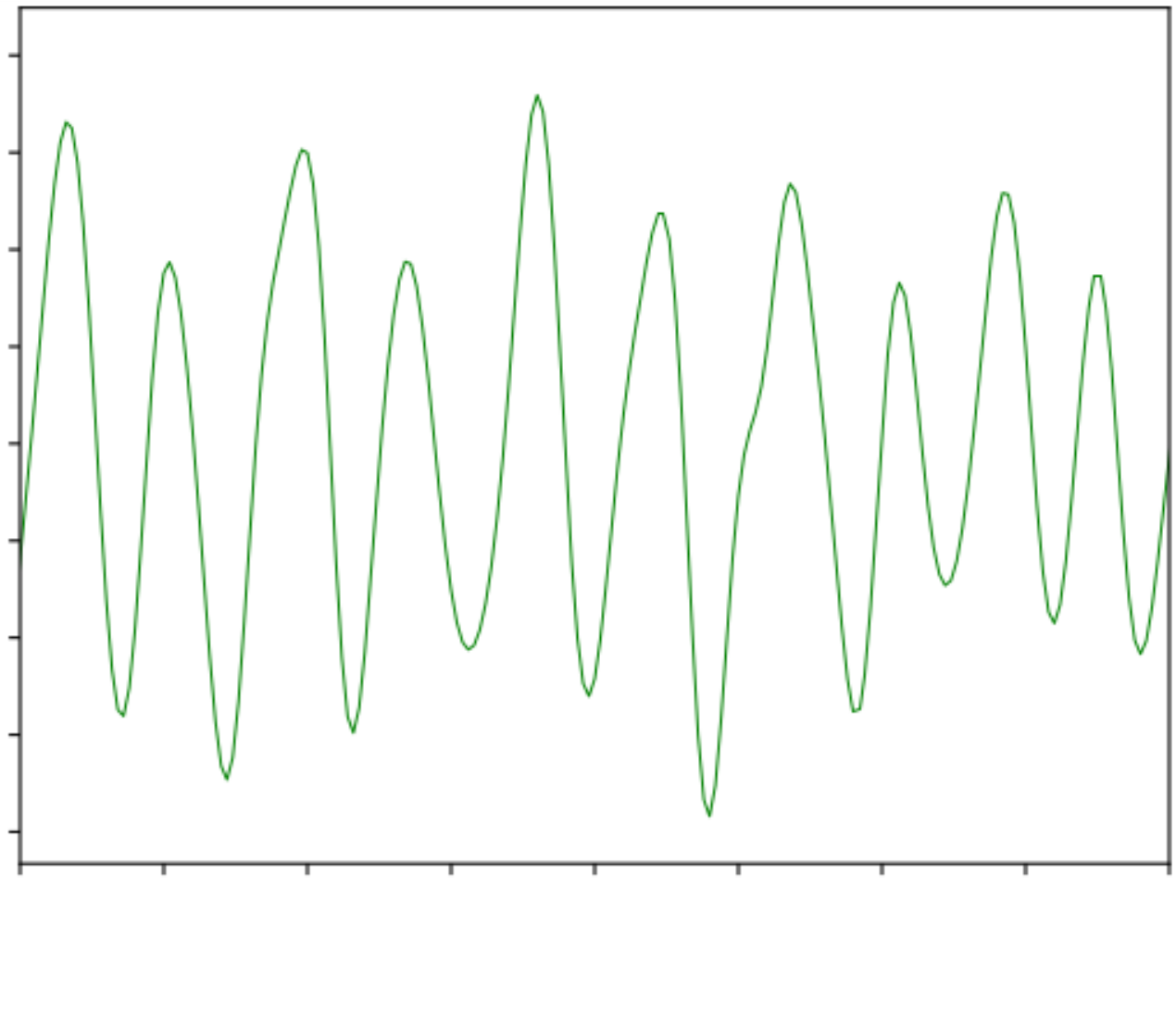}
            \put (-1, 7) {\small $12$}
            \put (11, 7) {\small $13$}
            \put (23, 7) {\small $14$}
            \put (35, 7) {\small $15$}
            \put (47, 7) {\small $16$}
            \put (59, 7) {\small $17$}
            \put (71, 7) {\small $18$}
            \put (84, 7) {\small $19$}
            \put (96, 7) {\small $20$}
            \put (42, 1) {\small Time $\left[ s \right]$}
            \put (-13, 15) {\small $-0.4$}
            \put (-13, 23) {\small $-0.3$}
            \put (-13, 31) {\small $-0.2$}
            \put (-13, 39) {\small $-0.1$}
            \put (-9, 48) {\small $0.0$}
            \put (-9, 56.5) {\small $0.1$}
            \put (-9, 65) {\small $0.2$}
            \put (-9, 73) {\small $0.3$}
            \put (-9, 81) {\small $0.4$}
            \put (-21, 30) {\rotatebox{90}{\small Signal $\frac{\mathcal{F}}{\mathcal{F}_{\mathrm{max}}}$ $ \left[ \cdot \right]$}}
        \end{overpic}}
    \caption{\small Comparison, for a given pixel, between raw signal (a) and filtered signal (b), for the same $8\ s$ of acquisition.}
    \label{fig:Confronto_segnale1}
\end{figure}

\begin{figure}[h!]
    \centering
    \begin{overpic}[width=0.9\textwidth,trim={2.2cm 2.4cm 2.2cm 2.2cm},clip]{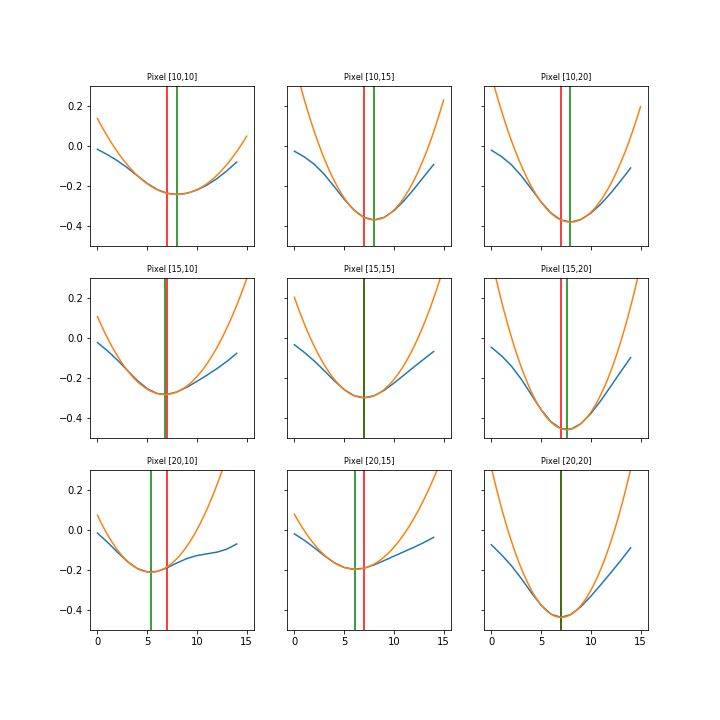}
    \end{overpic}
    \caption{\small Neuronal activity evaluated as $\frac{\mathcal{F}}{\mathcal{F}_{max}}$ in different macro-pixels, for $15$ frames (corresponding to $15\times40\ ms = 600\ ms$ of recordings). The miniplot graphic layout reproduces the pixels' spatial location on the hemisphere. In this visualization, the pixel in the centre is assumed as a reference; the signal is in blue, the minimum is fitted with a parabola (in orange), the green vertical line indicates the position of the minimum (from the vertex of the parabola). In the other plots, the red vertical line shows the time shift of the minimum with respect to the central pixel.}
    \label{fig:multiplot}
\end{figure}

The minima of the fluorescence signal are taken as markers of the passage of waves. By plotting the signal in a given time window for different channels (as in Figure \ref{fig:multiplot}), there is a visual confirmation that the neuronal activity occurring in the frequency band selected by the filter has a non-trivial spatio-temporal correlation. With the analysis of minima presented in the previous Section, we aim at disentangling the dynamics of the slow waves.
For each channel (\ie pixel), the collection of transition times is obtained as the vertex of the parabolic fit of the minima; 
the interpolation allows to reconstruct the onset of the upward transition overcoming the constraint of the temporal discretization in steps of $40\ ms$ due to to sampling frequency.
Representing the dynamics of the minima as in Figure \ref{fig:propagazione_onda}, the wave propagation phenomenon appears evident.
In addition, with the collection of transition times a \textit{raster plot} (Figure \ref{fig:raster}) can be produced. It offers a different perspective on the pixel activity: for each channel, the transition times are plotted as time passes. Here, the presence of a collective and periodic, yet not stereotyped, activity stands out. 
As already pointed out, it is notable that this plot is exactly the same (besides the number of channels) that can be obtained, at this step of processing, feeding the analysis pipeline with electrophysiological data collected via multi-electrode arrays. 


\begin{figure}
    \centering
    \subfloat[$t = 6,52\ s$]{
        \begin{overpic}[width=.46\textwidth]{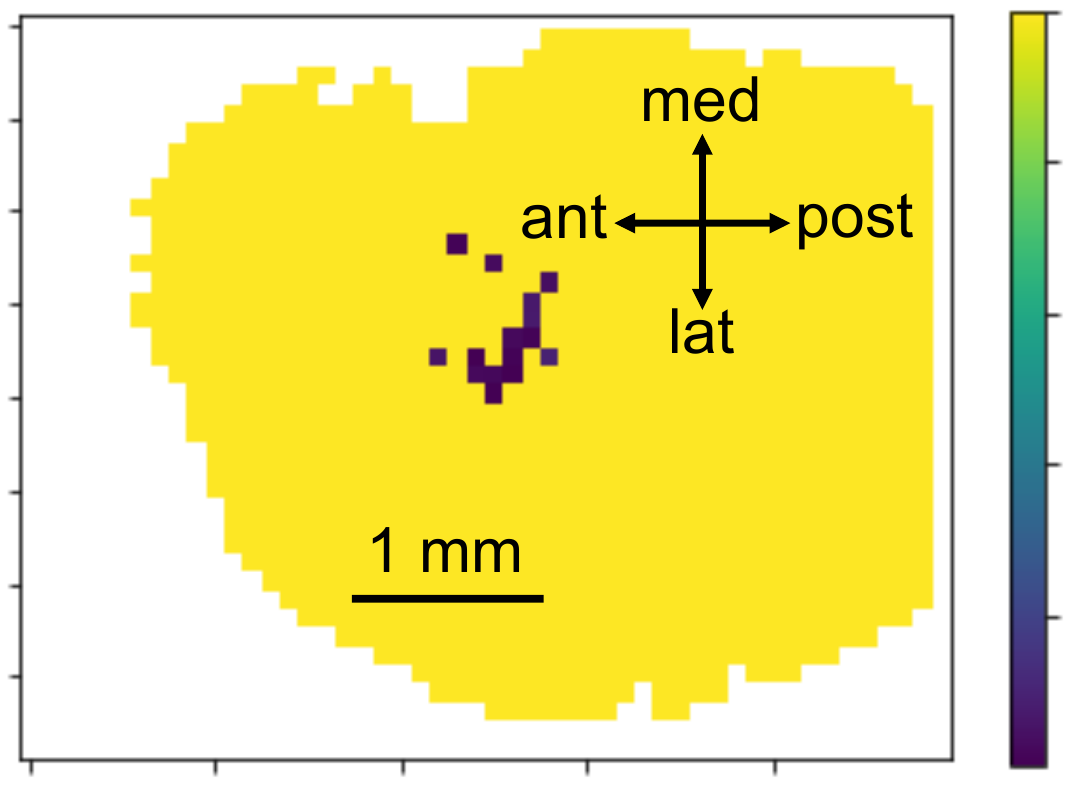}
            \put (91.5, 74) {\small [ms]}
            \put (102, 70) {\small $80$}
            \put (101, 57) {\small $64$}
            \put (101, 42) {\small $48$}
            \put (101, 28) {\small $32$}
            \put (101, 14) {\small $16$}
            \put (101, 1) {\small $0$}

            
    \end{overpic}} \qquad
    \subfloat[$t = 6,56s$]{
        \begin{overpic}[width=.46\textwidth]{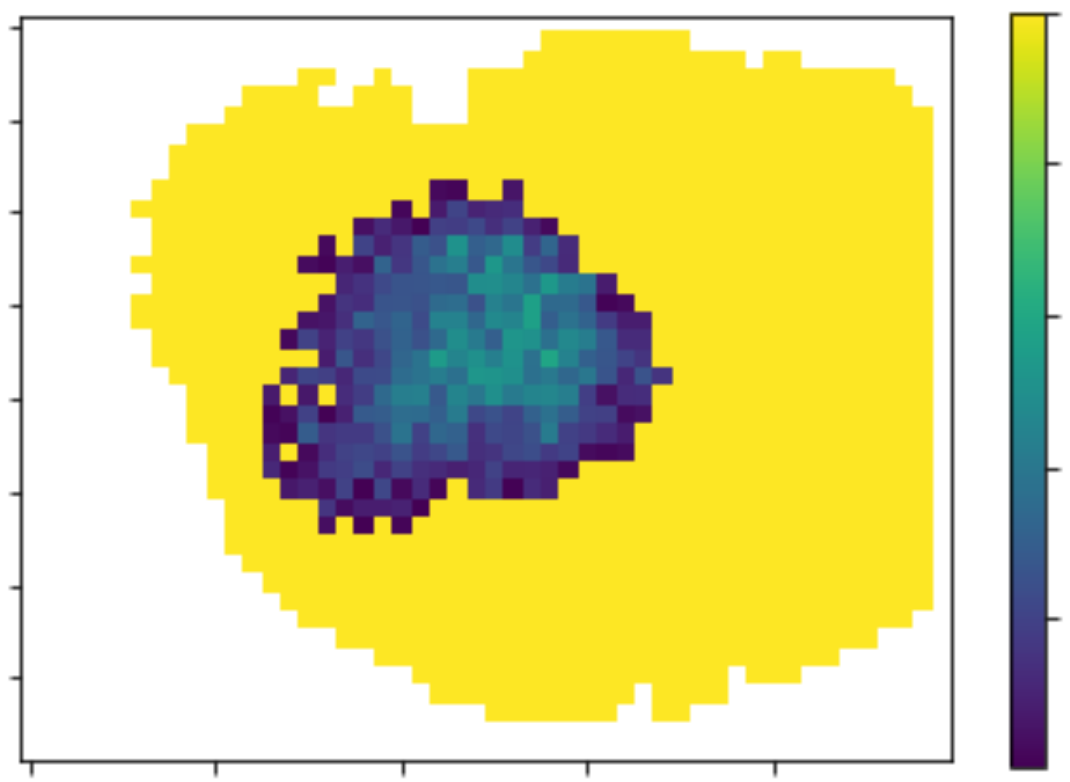}
            \put (91.5, 74) {\small [ms]}
            \put (102, 70) {\small $80$}
            \put (101, 57) {\small $64$}
            \put (101, 42) {\small $48$}
            \put (101, 28) {\small $32$}
            \put (101, 14) {\small $16$}
            \put (101, 1) {\small $0$}
        \end{overpic}} \\
    \subfloat[$t = 6,60\ s$]{
        \begin{overpic}[width=.46\textwidth]{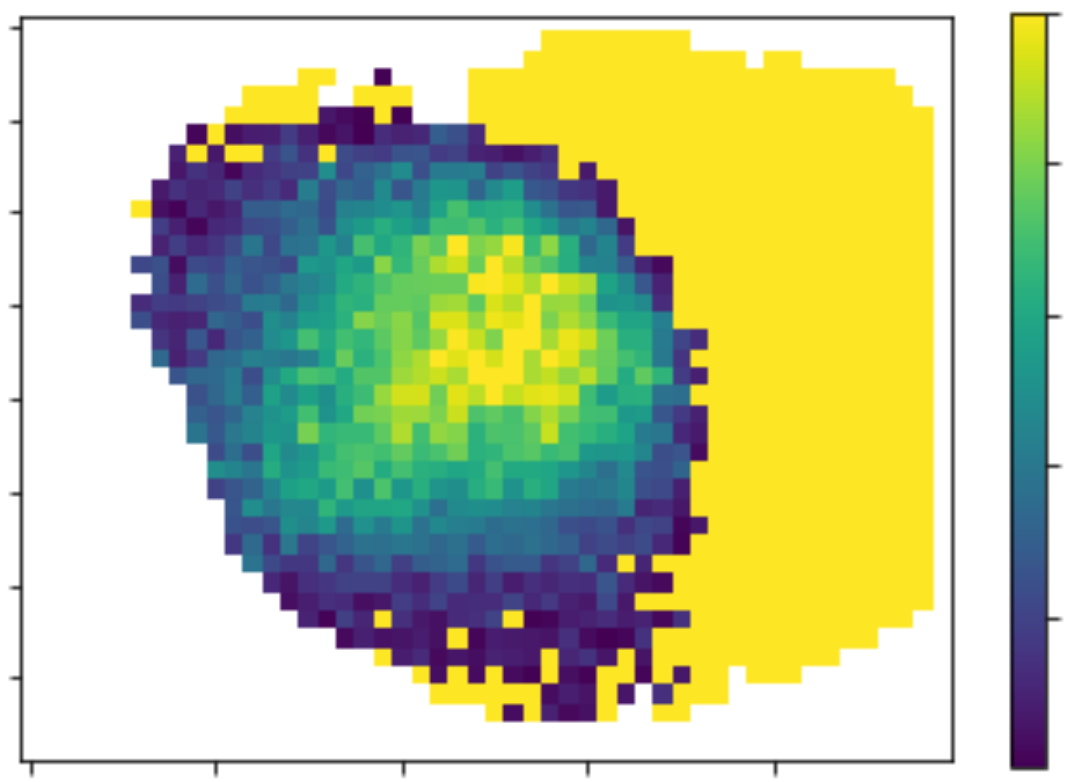}
            \put (91.5, 74) {\small [ms]}
            \put (102, 70) {\small $80$}
            \put (101, 57) {\small $64$}
            \put (101, 42) {\small $48$}
            \put (101, 28) {\small $32$}
            \put (101, 14) {\small $16$}
            \put (101, 1) {\small $0$}
        \end{overpic}} \qquad \ 
    \subfloat[$t = 6,64\ s$]{
        \begin{overpic}[width=.46\textwidth]{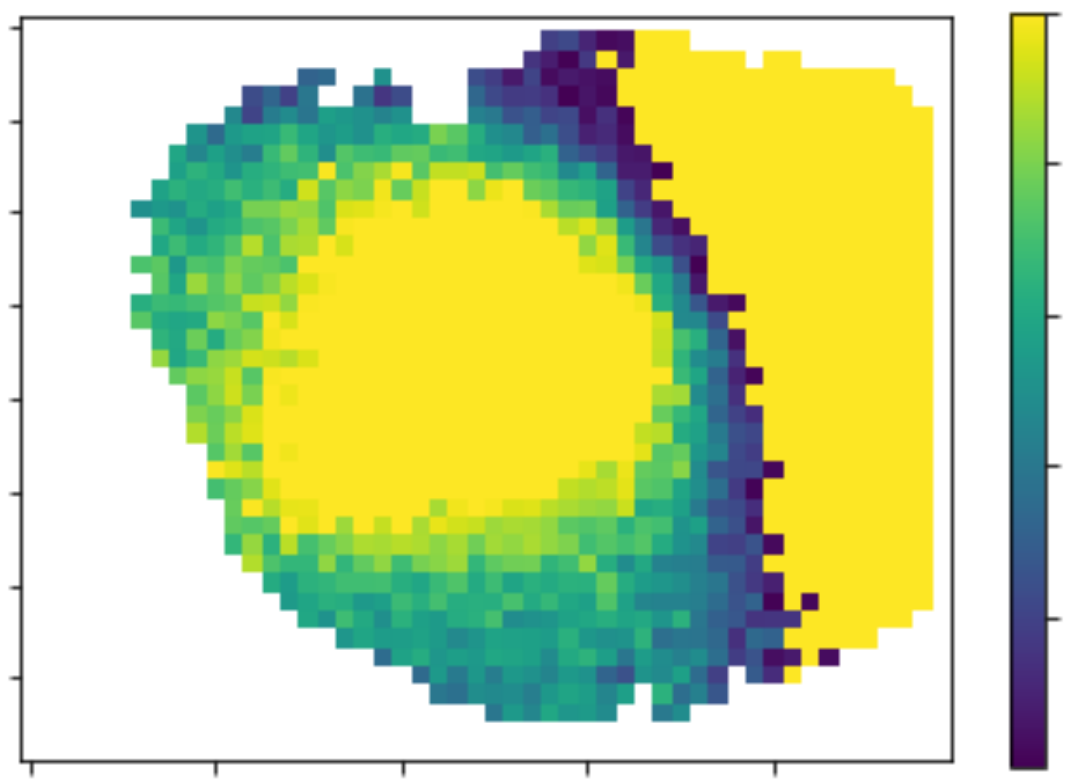}
            \put (91.5, 74) {\small [ms]}
            \put (102, 70) {\small $80$}
            \put (101, 57) {\small $64$}
            \put (101, 42) {\small $48$}
            \put (101, 28) {\small $32$}
            \put (101, 14) {\small $16$}
            \put (101, 1) {\small $0$}
        \end{overpic}} \\
    \subfloat[$t = 6,68\ s$]{
        \begin{overpic}[width=.46\textwidth]{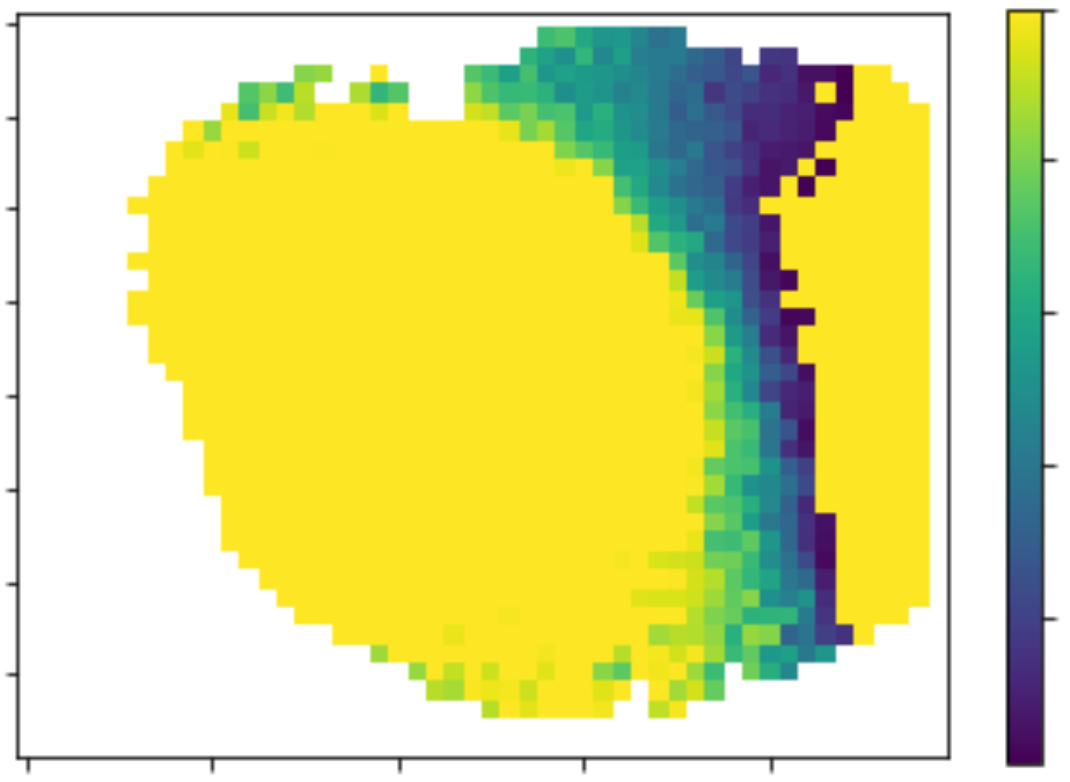}
            \put (91.5, 74) {\small [ms]}
            \put (102, 70) {\small $80$}
            \put (101, 57) {\small $64$}
            \put (101, 42) {\small $48$}
            \put (101, 28) {\small $32$}
            \put (101, 14) {\small $16$}
            \put (101, 1) {\small $0$}
        \end{overpic}} \qquad \ 
    \subfloat[$t = 6,72s$]{
        \begin{overpic}[width=.46\textwidth]{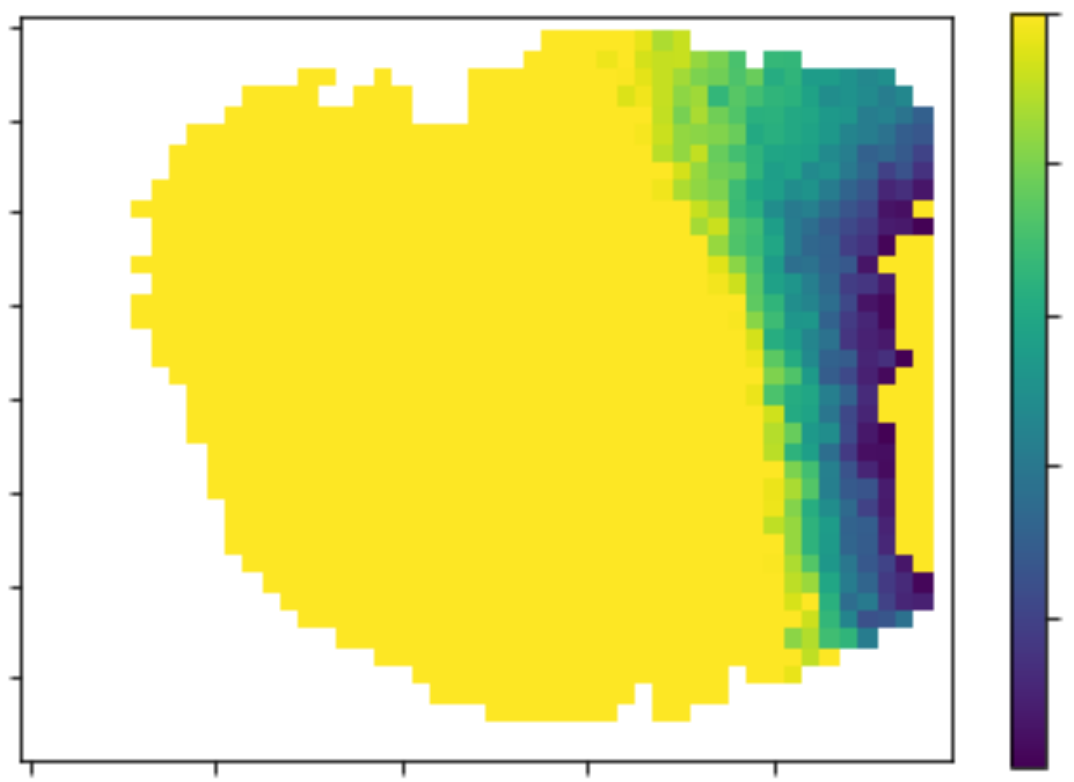}
            \put (91.5, 74) {\small [ms]}
            \put (102, 70) {\small $80$}
            \put (101, 57) {\small $64$}
            \put (101, 42) {\small $48$}
            \put (101, 28) {\small $32$}
            \put (101, 14) {\small $16$}
            \put (101, 1) {\small $0$}
        \end{overpic}}
    \caption{\small Series of images -- spaced by $40\ ms$ time step -- showing the propagation front of the minima. The global and correlated wave activity is evident. The pixels activated in the $80\ ms$ preceding the time shown below each image are illuminated. The color represents the time elapsed from the up-ward transition: it goes from the dark blue for the pixels that have just turned on, to the green for those that have been activated almost $80\ ms$ before the image capture time, up to the yellow for all the pixels whose neuronal population has not transited in the last $80\ ms$.}
    \label{fig:propagazione_onda}
\end{figure}
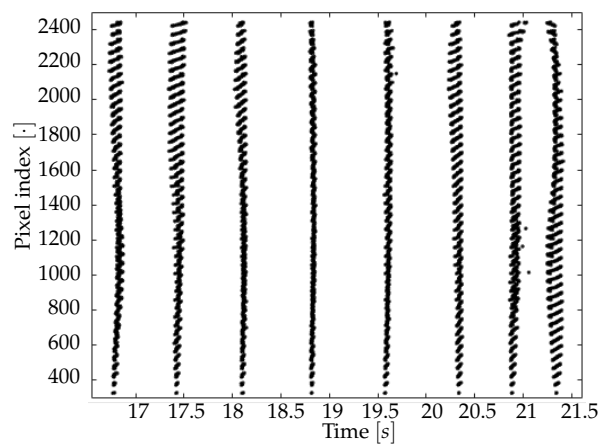
\begin{figure}[h!]
    \centering
    \begin{overpic}[width=0.42\textwidth]{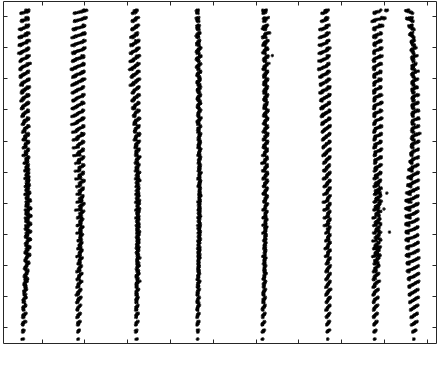}
        \footnotesize \put (8, 6) {$17$}
        \footnotesize\put (16, 6) {$17.5$}
        \footnotesize\put (27, 6) {$18$}
        \footnotesize\put (36, 6) {$18.5$}
        \footnotesize\put (47, 6) {$19$}
        \footnotesize\put (55, 6) {$19.5$}
        \footnotesize\put (67, 6) {$20$}
        \footnotesize\put (74.5, 6) {$20.5$}
        \footnotesize\put (86, 6) {$21$}
        \footnotesize\put (95, 6) {$21.5$}
        \footnotesize\put (47, 2) {Time $\left[ s \right]$}
        \footnotesize\put (-8, 13) {$400$}
        \footnotesize\put (-8, 20) {$600$}
        \footnotesize\put (-8, 27) {$800$}
        \footnotesize\put (-10, 34) {$1000$}
        \footnotesize\put (-10, 41) {$1200$}
        \footnotesize\put (-10, 48) {$1400$}
        \footnotesize\put (-10, 55) {$1600$}
        \footnotesize\put (-10, 62) {$1800$}
        \footnotesize\put (-10, 70) {$2000$}
        \footnotesize\put (-10, 77) {$2200$}
        \footnotesize\put (-10, 84) {$2400$}
        \footnotesize\put (-15, 40) {\rotatebox{90}{Pixel index $\left[ \cdot \right]$}}
    \end{overpic}
    \caption{\small Raster plot of a time interval for a given data set. The presence of a global and repeated activity of the system is clear, although each signal is characterized by different dynamics, reflecting differences in the propagation patterns.}
    \label{fig:raster}
\end{figure}

The WaveHunt algorithm, described in Section \ref{subsec:DataAnalysisPipeline}, splits the collection of transition times into global waves. Once this subdivision is completed and the trigger times of the waves are known, everything is on the table for moving to the last part of the analysis, and the quantitative measurements (Excitability, Origin Points and Velocity) can be taken; these will be discussed in Section \ref{sec:Discussion}.

\subsection{Analysis of the Simulation}
\label{subsec:SimulationAnalysis}

The purpose of the simulation is to generate a synthetic signal comparable with experimental data through the application of the same analysis procedures. This comparison aims at validating both the discussed analysis pipeline and the theoretical assumptions underlying the Toy Model.

\begin{figure}[h!]
    \centering
    \captionsetup[subfloat]{position=top}
    \subfloat[\footnotesize Single neuron firing rate]{
        \label{fig:NeuronSignalFiringRate}
        \hspace{0.5cm}
        \begin{overpic}[width=.43\textwidth,trim={1.8cm 1.575cm 0 0},clip]{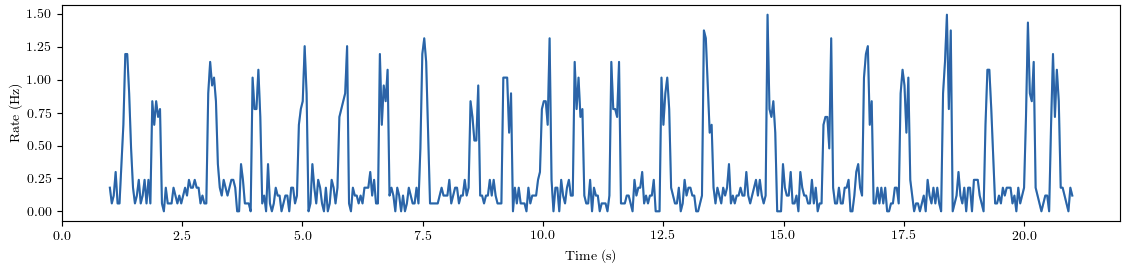}
            \put (-1, -4) {\footnotesize $0.0$}
            \put (10, -4) {\footnotesize $2.5$}
            \put (21, -4) {\footnotesize $5.0$}
            \put (32, -4) {\footnotesize $7.5$}
            \put (42, -4) {\footnotesize $10.0$}
            \put (53, -4) {\footnotesize $12.5$}
            \put (64, -4) {\footnotesize $15.0$}
            \put (75, -4) {\footnotesize $17.5$}
            \put (86, -4) {\footnotesize $20.0$}
            \put (40, -9) {\footnotesize Time $\left[ s \right]$}
            \put (-6, 0) {\footnotesize $0.0$}
            \put (-6, 6) {\footnotesize $0.5$}
            \put (-6, 12) {\footnotesize $1.0$}
            \put (-6, 18) {\footnotesize $1.5$}
            \put (-12, 3) {\rotatebox{90}{\footnotesize Rate $\left[ Hz \right]$}}
        \end{overpic}} \qquad \quad 
    \subfloat[\footnotesize Convolved signal (Lognormal kernel)]{
        \label{fig:NeuronSignal-Convolved}
        \hspace{0.1cm}
        \begin{overpic}[width=.43\textwidth,trim={2.4cm 1.475cm 0 0},clip]{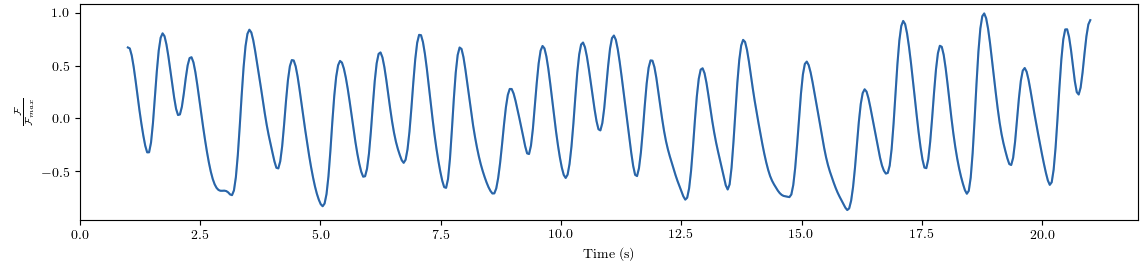}
            \put (-1, -4) {\footnotesize $0.0$}
            \put (10, -4) {\footnotesize $2.5$}
            \put (21, -4) {\footnotesize $5.0$}
            \put (32, -4) {\footnotesize $7.5$}
            \put (42, -4) {\footnotesize $10.0$}
            \put (53, -4) {\footnotesize $12.5$}
            \put (64, -4) {\footnotesize $15.0$}
            \put (75, -4) {\footnotesize $17.5$}
            \put (86, -4) {\footnotesize $20.0$}
            \put (40, -9) {\footnotesize Time $\left[ s \right]$}
            \put (-8, 4) {\scriptsize $-0.5$}
            \put (-6, 9) {\scriptsize $0.0$}
            \put (-6, 14) {\scriptsize $0.5$}
            \put (-6, 19) {\scriptsize $1.0$}    
            \put (-16, 3) {\rotatebox{90}{\footnotesize$\frac{\mathcal{F}}{\mathcal{F}_{\mathrm{max}}}$ $ \left[ \cdot \right]$}}
        \end{overpic}}
        \vspace{0.5cm}
    \caption{\small \textbf{(a)} Firing rate generated by a single \textit{Neuron} during a simulation. \textbf{(b)} Signal obtained after the convolution with a Lognormal kernel of parameters $ \mu = 2.5$ and $ \sigma = 0.5$, and including the discard of the transient time, the subtraction of the mean signal and the normalization to the maximum value.}
    \label{fig:Neuron_signal}
\end{figure}
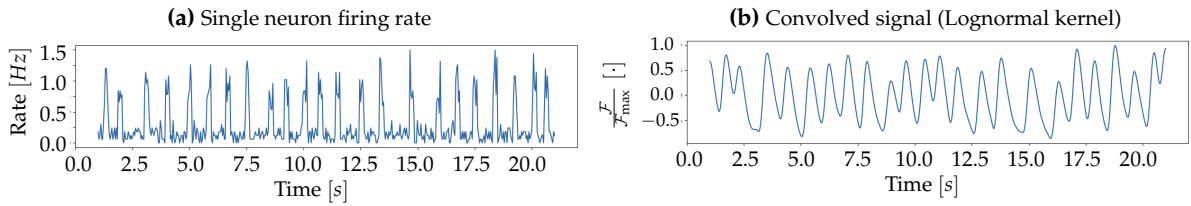

The signal is produced in successive steps. At the most basic level, every single \textit{Neuron} produces its own Poissonian signal with a firing rate that depends on the state of the \textit{Neuron}. This signal is then summed to that generated by the other \textit{Neurons} belonging to the same \textit{Pixel}, to form a \textit{Pixel} signal; the latter is then convolved with a kernel. An example of the Poissonian signal produced by a simulated  \textit{Neuron} is shown in Figure \ref{fig:NeuronSignalFiringRate}. 
The single \textit{Neuron} signal is flanked by the one obtained after the convolution with a Lognormal kernel (Figure \ref{fig:NeuronSignal-Convolved})\footnote{To be precise, since the convolution operation is distributive, it does not matter whether it is made before or after summing the \textit{Neuron} signals. Therefore, although in Figure \ref{fig:Neuron_signal} a convolved \textit{Neuron} signal is shown, its shape would be very similar to the one obtained applying the Lognormal kernel to the \textit{Pixel} signal.}.
The intensity value of the convolved signal is normalized to its maximum, in order to obtain an adimensional quantity; in analogy with the processing of the experimental data, the normalization is done after the subtraction of the constant non-informative component (\ie the average signal), in order to focus on the signal's fluctuations.
However, the very first action with simulated signals is to discard the initial part of the simulation; indeed, the convolution with a response function having a non-zero rise time induces a measurable delay between the Poisson signal of the single \textit{Neurons} and the brightness signal of the \textit{Pixel}, that would produce a low-frequency anomalous component in the spectral decomposition of the simulated signal. To overcome this effect, the transient time has been discarded, motivated by the fact that the experimental data are images of cortical activity, and a possible period of initial rising of the light signal due to the initial activation of the network is absent, or in any case not measurable. By eliminating the first second of the simulation (as shown in Figure \ref{fig:NeuronSignal-Convolved}), a spectral trend in accordance with the experimental counterpart is recovered for the simulated signal.

The parameters $\mu$ and $\sigma$ of the response function (kernel) have been chosen to obtain the highest agreement with the experimental trend of the GCaMP6 protein. 
We recall here the formula adopted in this work for the convolution kernel, specializing it with the parameters used for the simulation (listed in in Table \ref{tab:Tabella_Simulazione}); the presence of $ t_{\mathrm{step}} $, time step of the simulation, is necessary to make the time parameter adimensional:
\begin{equation}
    \begin{split}
        \mathrm{Lognormal} \left(\frac{t}{t_{\mathrm{step}}}; \mu, \sigma \right) = \frac{t_{\mathrm{step}}}{t} \frac{1}{\sqrt{2 \pi} \sigma} \exp \left( - \frac{\left( \ln \frac{t}{t_{\mathrm{step}}} - \mu \right)^2}{2 \sigma^2} \right).
        \label{eq:LogNormalAdopted}
    \end{split}
\end{equation}


\begin{table}[h!]
    \begin{center}
        \begin{tabular}{cp{7cm}c} \hline \vspace{-0.15cm} \\
            Parameter & \hspace{1cm} Description & Value [unit]\\ \vspace{-0.15cm} \\\hline \\
            $t_{\mathrm{end}}$ & Total time to simulate & Defined in input $\left[ s \right]$\\
            $t_{\mathrm{step}}$ & Simulation time step & $40\ \left[ms \right]$\\
            $\tau_{\mathrm{up}}$ & Neuron permanence time in active state & $200\ \left[ms \right]$\\
            $\frac{\mu_{\mathrm{up}}}{\mu_{\mathrm{down}}}$ & Ratio of firing rates (active and idle states) & $5 \left[ \cdot \right]$ \\
            $N_{\mathrm{pixel}}$ & Neuron number for each Pixel & $N \simeq \mathcal{N} \left(\mu = 10, \sigma = 2 \right)\ \left[ \cdot \right]$\\
            $\mu$ & $\mu$ parameter of the LogNorm Kernel & $2.2\ \left[ \cdot \right]$\\
            $\sigma$ & $\sigma$ parameter of the LogNorm Kernel & $0.91\ \left[ \cdot \right]$\\
            \vspace{0.1cm} \\ \hline \\[0.1cm]
        \end{tabular}
        \caption{\small Simulation parameters}
        \label{tab:Tabella_Simulazione}
    \end{center}
\end{table}
\subsection{Comparison between Experimental Samples}
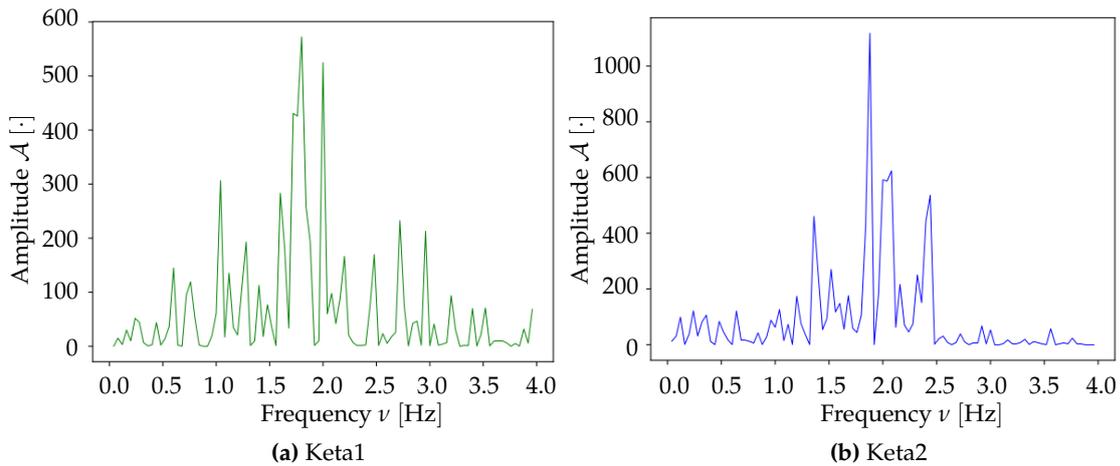
\begin{figure}[h!]
    \centering
    \subfloat[Keta1]{\label{fig:Confronto_spettro_Keta1}
        \begin{overpic}[width = 0.40\textwidth]{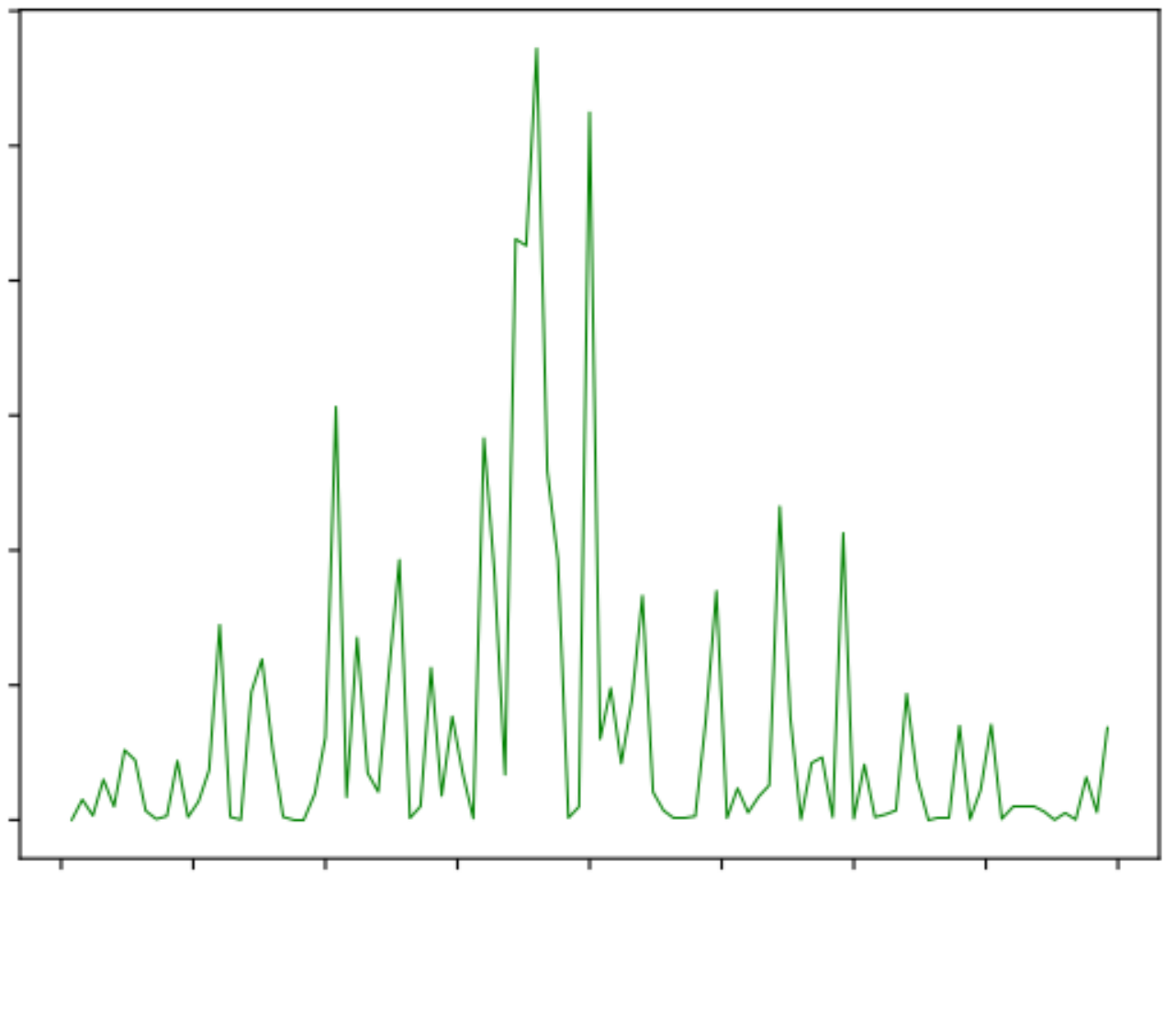}
            \put (3, 7) {\small $0.0$}
            \put (14, 7) {\small $0.5$}
            \put (25, 7) {\small $1.0$}
            \put (36, 7) {\small $1.5$} 
            \put (48, 7) {\small $2.0$}
            \put (59, 7) {\small $2.5$}
            \put (70, 7) {\small $3.0$}
            \put (81, 7) {\small $3.5$}
            \put (93, 7) {\small $4.0$}
            \put (37, 1) {\small Frequency $\nu$ $\left[\mathrm{Hz} \right]$}
            \put (-4, 15) {\small $0$}
            \put (-9, 27) {\small $100$}
            \put (-9, 38) {\small $200$}
            \put (-9, 50) {\small $300$}
            \put (-9, 61) {\small $400$}
            \put (-9, 73) {\small $500$}
            \put (-9, 85) {\small $600$}
            \put (-16, 30) {\rotatebox{90}{\small Amplitude $\mathcal{A}$ $\left[ \cdot \right]$}}
        \end{overpic}} \qquad
    \hspace{0.25cm}
    \subfloat[Keta2]{\label{fig:Confronto_spettro_Keta2}
        \begin{overpic}[width = 0.40\textwidth]{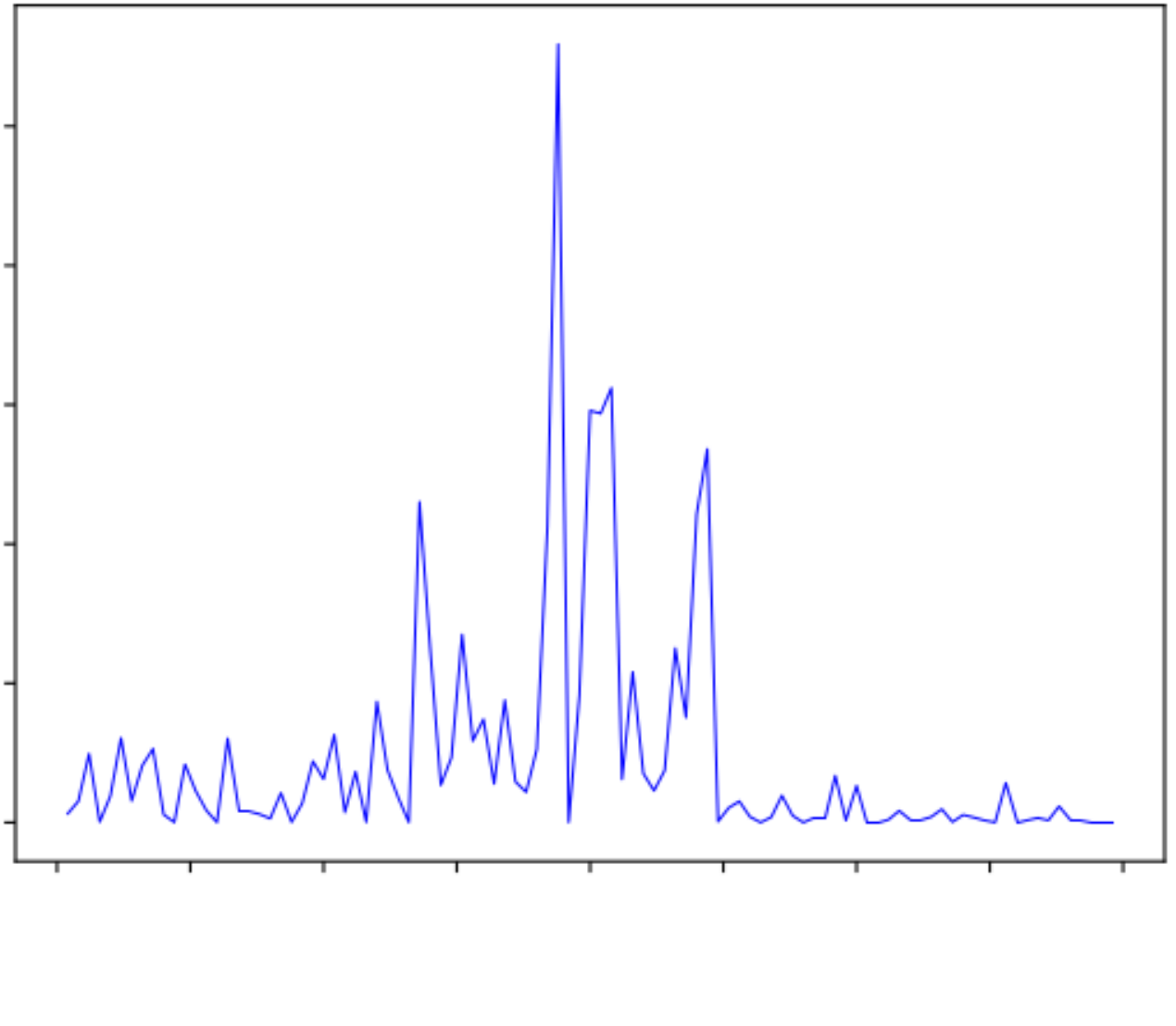}
            \put (3, 7) {\small $0.0$}
            \put (14, 7) {\small $0.5$}
            \put (25, 7) {\small $1.0$}
            \put (36, 7) {\small $1.5$}
            \put (48, 7) {\small $2.0$}
            \put (59, 7) {\small $2.5$}
            \put (70, 7) {\small $3.0$}
            \put (81, 7) {\small $3.5$}
            \put (93, 7) {\small $4.0$}
            \put (37, 1) {\small Frequency $\nu$ $\left[\mathrm{Hz} \right]$}
            \put (-4, 15) {\small $0$}
            \put (-9, 28) {\small $200$}
            \put (-9, 39) {\small $400$}
            \put (-9, 52) {\small $600$}
            \put (-9, 64) {\small $800$}
            \put (-11, 75) {\small $1000$}
            \put (-16, 30) {\rotatebox{90}{\small Amplitude $\mathcal{A}$ $\left[ \cdot \right]$}}
        \end{overpic}}	
    \caption{\small Comparison between the average frequency spectra of Keta1 \textbf{(a)} (same plot as in Figure \ref{fig:spettro_170110}) and Keta2 \textbf{(b)}. Both samples present a peak at similar frequency values in the delta band.}
    \label{fig:Confronto_Spettro}
\end{figure}

We compare the results from the two experimental datasets, acquired as presented in \ref{subsec:Methods-DAQ} and referred to as \emph{Keta1} and \emph{Keta2}.
In both mice, an oscillatory signal is observed in the delta waves frequency band. For the Keta1 mouse, the main peak in the frequency spectrum is found at $f_{\mathrm{max}} \simeq 1.7\ \mathrm{Hz}$ (Figure \ref{fig:Confronto_spettro_Keta1}); for the Keta2 mouse, the most intense frequency component is observed at $f_{\mathrm{max}} \simeq 1.9\ \mathrm{Hz}$ (Figure \ref{fig:Confronto_spettro_Keta2}); the two spectral activities appear similar in the comparison, suggesting that a similarity may extend to other features of the samples. 
%

\begin{figure}[h!]
    \centering
    \subfloat[Keta1]{\label{fig:WaveOrigin_Keta1}
        \begin{overpic}[width = .44\textwidth]{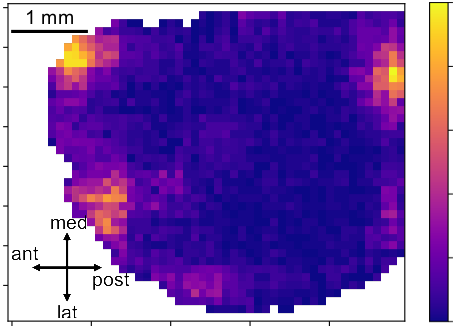}
            \put (102, 0) {\small $0$}
            \put (101, 14) {\small $0.2$}
            \put (101, 28) {\small $0.4$}
            \put (101, 42) {\small $0.6$}
            \put (101, 56) {\small $0.8$}
            \put (101, 70) {\small $1.0$}


        \end{overpic}} \qquad \quad
    \subfloat[Keta2]{\label{fig:WaveOrigin_Keta2}
        \begin{overpic}[width = .44\textwidth]{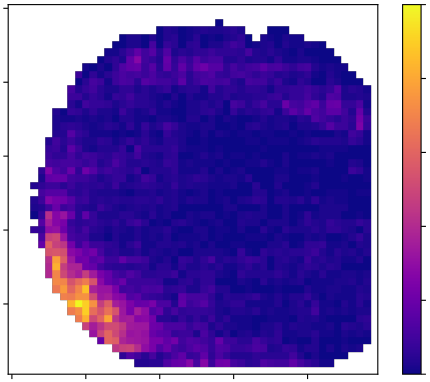}
            \put (102, 2) {\small $0$}
            \put (101, 18) {\small $0.2$}
            \put (101, 35) {\small $0.4$}
            \put (101, 53) {\small $0.6$}
            \put (101, 69) {\small $0.8$}
            \put (101, 86) {\small $1.0$}
            


        \end{overpic}}
    \caption{\small Maps of the point of origin of the experimental waves for Keta1 \textbf{(a)} and Keta2 \textbf{(b)}. The color code represents the number of times a single pixel has been involved in the birth of a global wave, normalized to the total number of waves in the collection. The ``birth set'' is defined as the first $N = 30$ pixels on which each wave passes; given the request of globality for the waves included in the collection, and taking into account the different numbers of informative pixels in the two datasets, the birth set constitutes at most the $ 3\% $ of the wave.}
\end{figure}
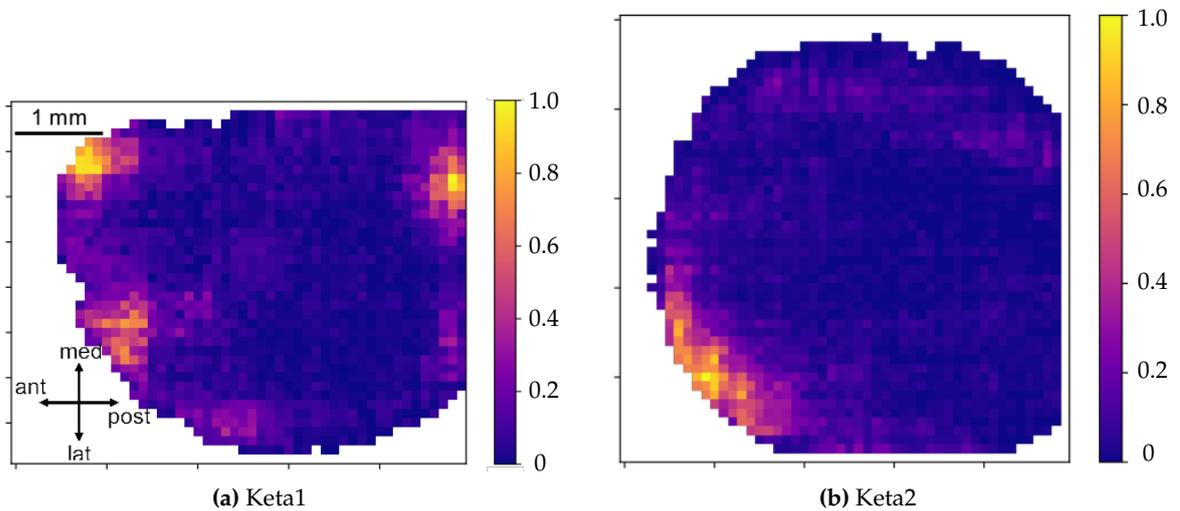

The comparison of the graphs with the points of origin of the waves (Figure \ref{fig:WaveOrigin_Keta1} and \ref{fig:WaveOrigin_Keta2}) shows an overall similarity in the activity observed in Keta1 and in Keta2. In both samples, in fact, there is a prevalence of waves that start from the prefrontal cortex, although in Keta1 there is a spot of origin also in the posterior cortex -- indicatively in the retrosplenial area -- and a wider distribution of the points of origin in the frontal cortex.
 
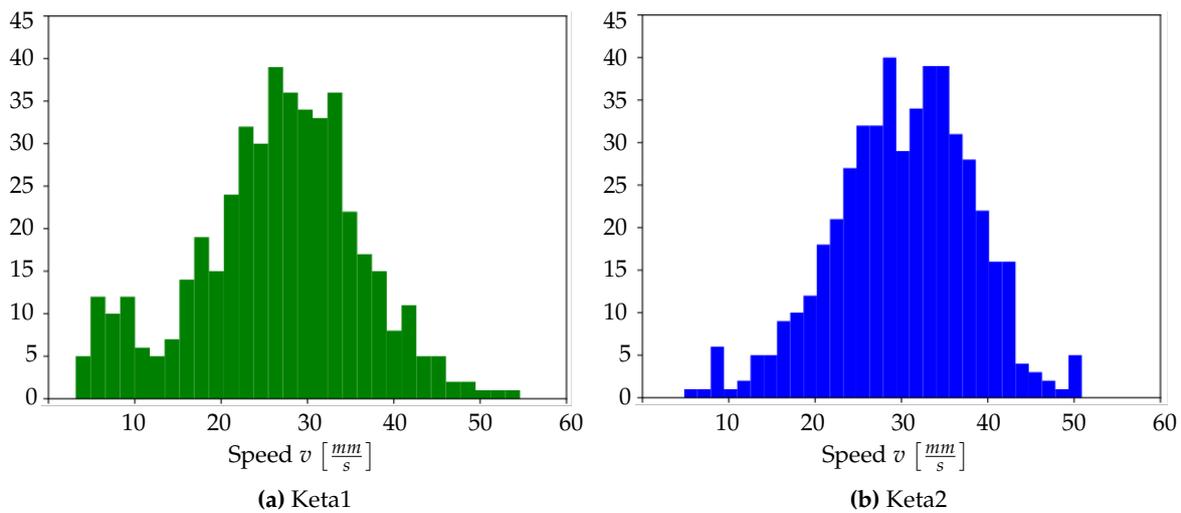
\begin{figure}[h!]
    \centering
    \subfloat[Keta1]{\label{fig:WaveVelocity_Keta1}
        \begin{overpic}[width=0.45\textwidth]{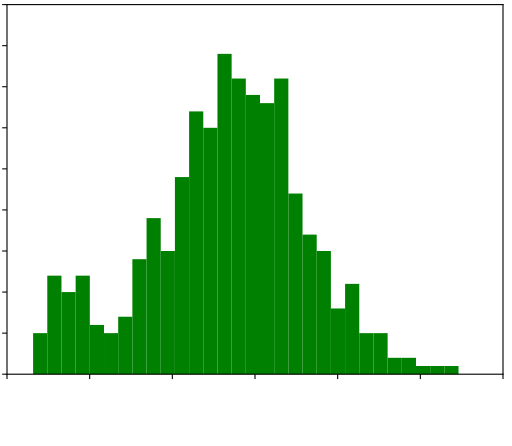}
            \put (15, 8) {\small $10$}
            \put (31, 8) {\small $20$}
            \put (48, 8) {\small $30$}
            \put (64, 8) {\small $40$}
            \put (80, 8) {\small $50$}
            \put (97, 8) {\small $60$}

            \put (35, 2) {\small Speed $v$ $\left[\frac{mm}{s} \right]$}
            \put (-3, 13) {\small $0$}
            \put (-3, 21) {\small $5$}
            \put (-6, 29) {\small $10$}
            \put (-6, 37) {\small $15$}
            \put (-6, 45) {\small $20$}
            \put (-6, 53) {\small $25$}
            \put (-6, 61) {\small $30$}
            \put (-6, 69) {\small $35$}
            \put (-6, 77) {\small $40$}
            \put (-6, 84) {\small $45$}
        \end{overpic}} \qquad
    \subfloat[Keta2]{\label{fig:WaveVelocity_Keta2}
        \begin{overpic}[width=0.45\textwidth]{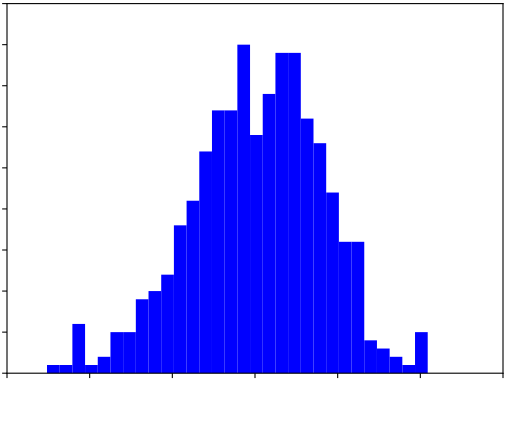}
            \put (15, 8) {\small $10$}
            \put (31, 8) {\small $20$}
            \put (48, 8) {\small $30$}
            \put (64, 8) {\small $40$}
            \put (80, 8) {\small $50$}
            \put (97, 8) {\small $60$}
            \put (35, 2) {\small Speed $v$ $\left[\frac{mm}{s} \right]$}
            \put (-3, 13) {\small $0$}
            \put (-3, 21) {\small $5$}
            \put (-6, 29) {\small $10$}
            \put (-6, 37) {\small $15$}
            \put (-6, 45) {\small $20$}
            \put (-6, 53) {\small $25$}
            \put (-6, 61) {\small $30$}
            \put (-6, 69) {\small $35$}
            \put (-6, 77) {\small $40$}
            \put (-6, 84) {\small $45$}
        \end{overpic}}
    \caption{\small Histogram of the wavefront speed for Keta1 \textbf{(a)} and Keta2 \textbf{(b)}.}
\end{figure}

Performing the study of the average wave propagation speed, we obtain the two histograms represented in Figure \ref{fig:WaveVelocity_Keta1} and \ref{fig:WaveVelocity_Keta2}. For Keta1, the average propagation speed is $\langle v \rangle_{\mathrm{Keta}_1} = \left( 26 \pm 8 \right)\ \frac{mm}{s}$; for Keta2, the average speed is $\langle v \rangle_{\mathrm{Keta}_2} = \left(30 \pm 8 \right)\ \frac{mm}{s}$.
Not only these results are consistent with each other, they are also in agreement with other propagation measurements found in the literature. For example, in \cite{Ruiz-Mejias2011:Slow-and-fast-rhythms}, a study carried out using electrophysiological techniques, the values $\langle v \rangle = \left(30.0 \pm 3.9 \right)$ $\frac{mm}{s}$ and  $\langle v \rangle = \left(23.4 \pm 2.1 \right)$ $\frac{mm}{s}$ are reported for recordings from the motor and the visual cortex, respectively, for a mouse anesthetized with a mix of ketamine (75 mg / kg) and medetomidine (1 mg / kg).

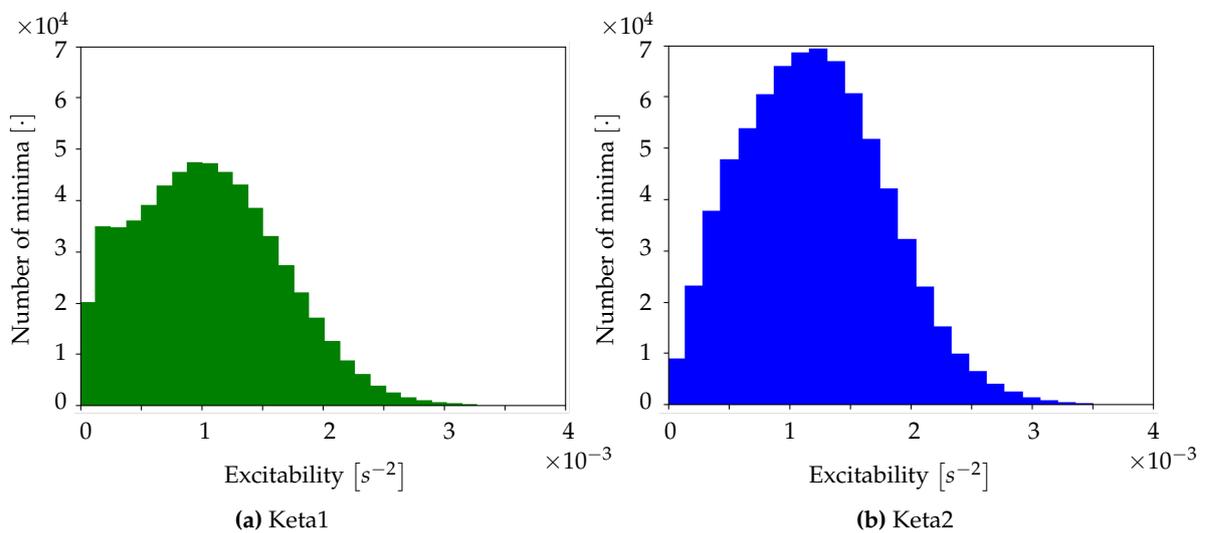
\begin{figure}[h!]
    \subfloat[Keta1]{\label{fig:Confronto_Ecc_hist_K1}\hspace{1cm}
        \begin{overpic}[width=0.42\textwidth]{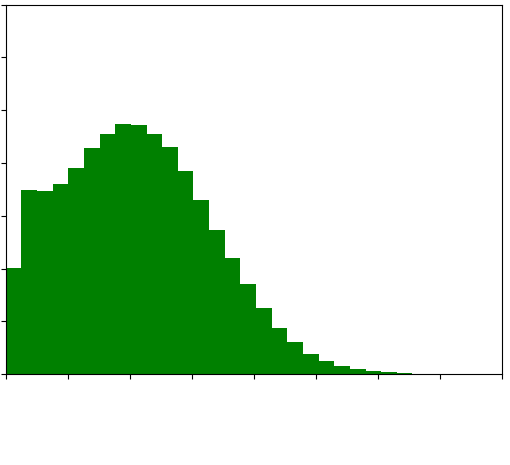}
            \put (1, 11) {\small $0$}
            \put (25, 11) {\small $1$}
            \put (50, 11) {\small $2$}
            \put (74, 11) {\small $3$}
            \put (98, 11) {\small $4$}
            \put (94, 5) {\small $\times 10^{-3}$}
            \put (30, 2) {\small Excitability $\left[s^{-2} \right]$}
            \put (-4, 17) {\small $0$}
            \put (-4, 27) {\small $1$}
            \put (-4, 37) {\small $2$}
            \put (-4, 48) {\small $3$}
            \put (-4, 59) {\small $4$}
            \put (-4, 68) {\small $5$}
            \put (-4, 78) {\small $6$}
            \put (-4, 88) {\small $7$}
            \put (-12, 93) {\small $\times 10^{4}$}
            \put (-13, 30) {\rotatebox{90}{\small Number of minima $\left[ \cdot \right]$}}
        \end{overpic}} \qquad \quad
    \subfloat[Keta2]{\label{fig:Confronto_Ecc_hist_K2}
        \begin{overpic}[width=0.42\textwidth]{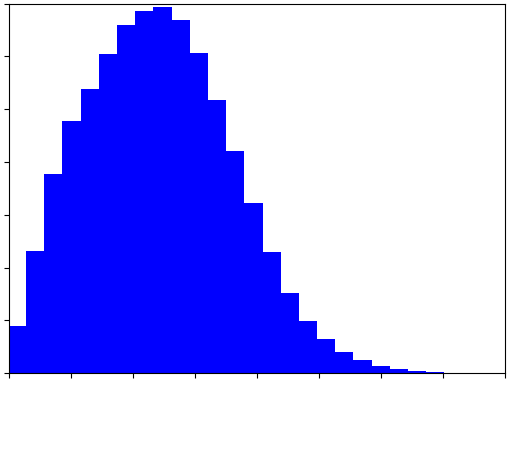}
            \put (1, 11) {\small $0$}
            \put (25, 11) {\small $1$}
            \put (50, 11) {\small $2$}
            \put (74, 11) {\small $3$}
            \put (98, 11) {\small $4$}
            \put (94, 5) {\small $\times 10^{-3}$}
            \put (30, 2) {\small Excitability $\left[s^{-2} \right]$}
            \put (-4, 17) {\small $0$}
            \put (-4, 27) {\small $1$}
            \put (-4, 37) {\small $2$}
            \put (-4, 48) {\small $3$}
            \put (-4, 59) {\small $4$}
            \put (-4, 68) {\small $5$}
            \put (-4, 78) {\small $6$}
            \put (-4, 88) {\small $7$}
            \put (-12, 93) {\small $\times 10^{4}$}
            \put (-13, 30) {\rotatebox{90}{\small Number of minima $\left[ \cdot \right]$}}
        \end{overpic}}
    \caption{\small Histogram of excitability for Keta1 \textbf{(a)} and Keta2 \textbf{(b)}.}
    \label{fig:ExcitabilityHistos}
\end{figure}

The coherence between the neuronal activity of Keta1 and Keta2 is evident also looking at the comparison of the excitability histograms (Figure \ref{fig:Confronto_Ecc_hist_K1} and \ref{fig:Confronto_Ecc_hist_K2}). For Keta1, the average excitability is $\varepsilon_{\mathrm{Keta}_1} = \left( 1.0 \pm 0.6\right) \times 10^{-3}\ s^{-2}$, whereas for Keta2 the average excitability is $\varepsilon_{\mathrm{Keta}_2} = \left( 1.5 \pm 0.5\right) \times 10^2\ s^{-2}$; these two values are compatible with each other. The histograms also present a characteristic pattern with most of the neuronal population sharing excitability close to the unit, but with an evident asymmetry that denotes the presence of non-negligible high excitability tails.

\begin{figure}[h!]
    \hspace{0cm}
    \subfloat[Keta1]{\label{fig:Mappe_ecc_Keta1}
        \begin{overpic}[width=.42\textwidth]{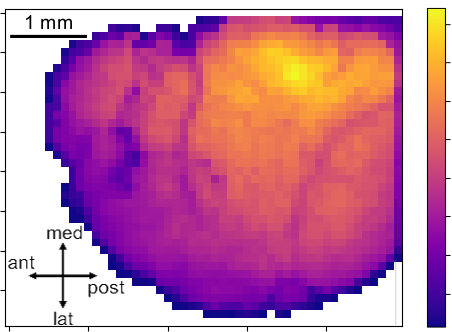}
            \put (101, 7) {\small $0.25$}
            \put (101, 16) {\small $0.50$}
            \put (101, 24) {\small $0.75$}
            \put (101, 33) {\small $1.00$}
            \put (101, 42) {\small $1.25$}
            \put (101, 51) {\small $1.50$}
            \put (101, 59) {\small $1.75$}
            \put (101, 67) {\small $2.00$}
            \put (101, 0) {$\left[ s^{-2} \right]$}
            \put (101, 75) {\small $\times 10^{-3}$}
            

            
        \end{overpic}} \qquad
    \hspace{0.8cm}
    \subfloat[Keta2]{\label{fig:Mappe_ecc_Keta2}
        \begin{overpic}[width=.42\textwidth]{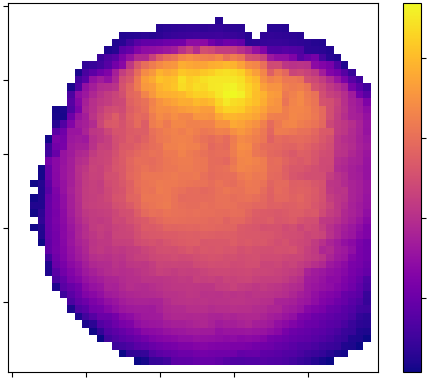}
            \put (101, 7) {\small $0.25$}
            \put (101, 18) {\small $0.50$}
            \put (101, 27) {\small $0.75$}
            \put (101, 37) {\small $1.00$}
            \put (101, 46) {\small $1.25$}
            \put (101, 55) {\small $1.50$}
            \put (101, 65) {\small $1.75$}
            \put (101, 74) {\small $2.00$}
            \put (101, 0) {$\left[ s^{-2} \right]$}
            \put (101, 85) {\small $\times 10^{-3}$}
            

        \end{overpic}}
    \caption{\small Maps of the cortex showing the average excitability per pixel for Keta1 \textbf{(a)} and Keta2 \textbf{(b)}.}
\end{figure}
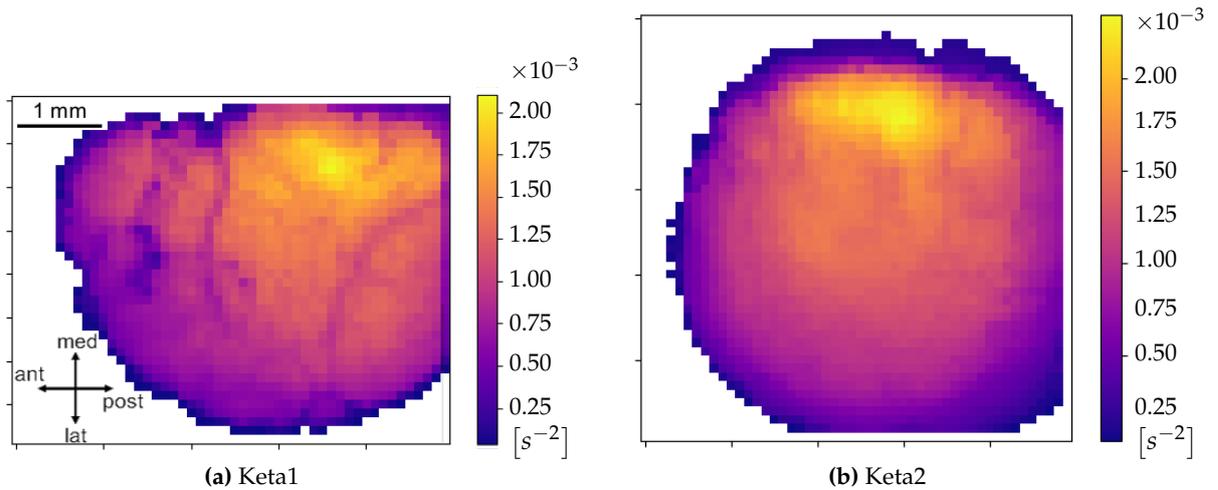

An additional graphical representation of the excitability data is shown in Figure \ref{fig:Mappe_ecc_Keta1} and \ref{fig:Mappe_ecc_Keta2} where the spatial dependence of this parameter is made explicit. In this case, for each channel, the mean excitability is computed, allowing to monitor how the mean excitability varies in the different cortical regions. Both Keta1 and Keta2 show a peak of excitability in the central part of the inspected cortex. 
%

\subsection{Comparison between Experimental and Simulated Data}
By using the experimental collection of transition times as a trigger for the activation of \textit{Pixels} in the Toy Model, it is possible to verify whether or not the simulation can reproduce the acquired signals;
the comparison has been made over a time frame of $40\ s$ of recordings (\ie on a single dataset).
As can be seen in Figure \ref{fig:Confronto_spettro_Sim} and \ref{fig:Confronto_spettro_Keta2_sim}, the agreement between the average frequency spectrum of the experimental data and the one obtained from the Toy Model is very good\footnote{As already discussed in section \ref{subsec:SimulationAnalysis}, the frequency spectrum computed on the simulated activity has been obtained by eliminating the first second of simulation, a choice made in order to avoid that the activation period of the Toy Model would spoil the spectrum, introducing a large low frequency component not present once the simulation has reached its effective working regime.}.
Moreover, as illustrated in Figure \ref{fig:Confronto_segnale}, the simulated signal shows a very convincing agreement with the one recorded \textit{in vivo} from the mouse's cortex \footnote{The signal produced by the simulation has been shifted by $ 2.93\ ms$ to obtain Figure \ref{fig:Confronto_segnale}; this value follows by calculating the correlation function between the two signals.}.
Results in Figures \ref{fig:Confronto_Spettro_Sim_Keta2}--\ref{fig:Confronto_segnale}  validate our hypothesis about the transfer function adopted to represent the fluorescence response of the GCaMP6f protein, and also indicate that the simplistic assumptions we made for the Toy Model are somehow grounded, at least enough to suggest a confirmation that it is possible to detect the upward transitions using optical techniques, overcoming the differences in the time scales of the underlying phenomena, and the constraints of the sampling frequency. 

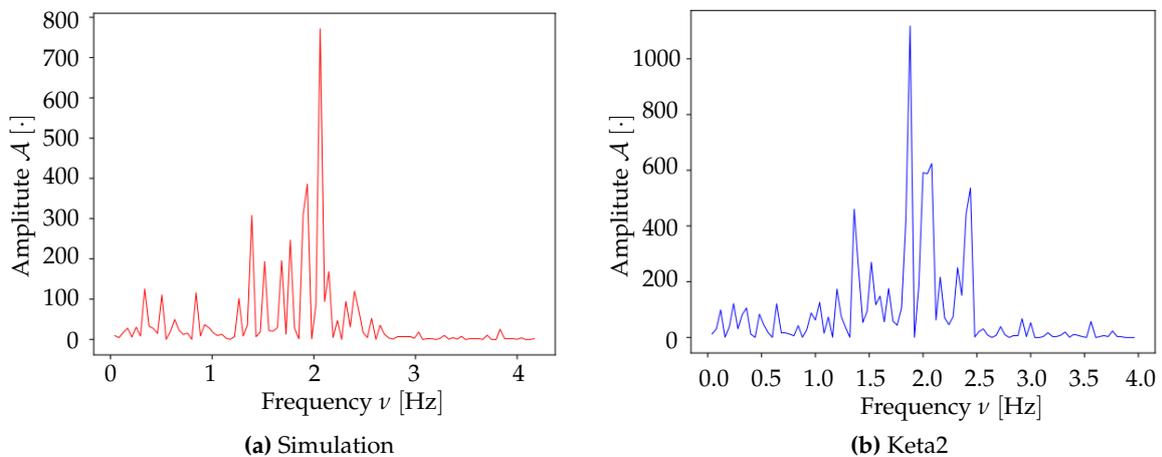
\begin{figure}[h!]
    \hspace{1cm}\subfloat[Simulation]{\label{fig:Confronto_spettro_Sim}
        \begin{overpic}[width=.4\textwidth]{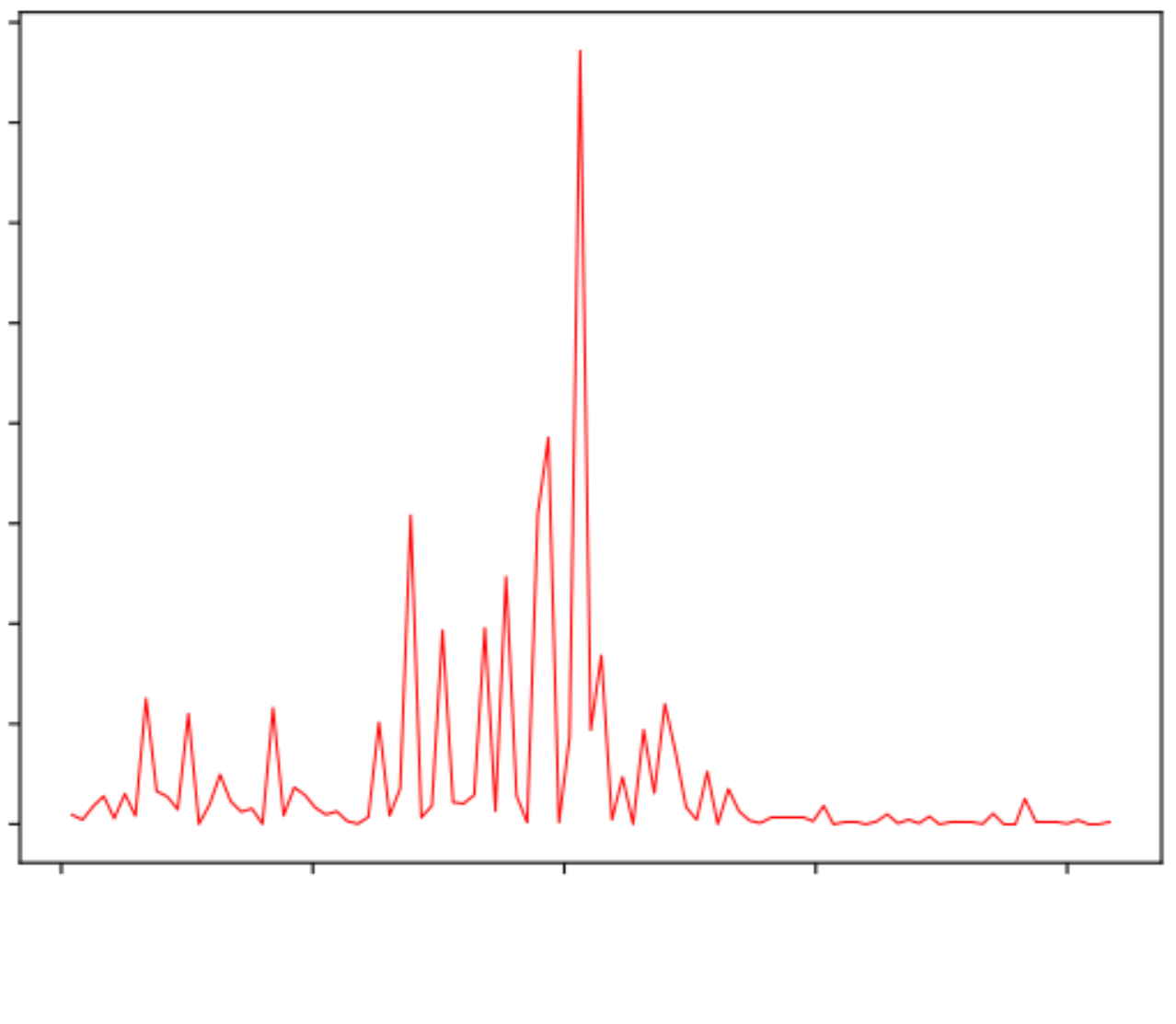}
            \put (4, 8) {\small $0$}
            \put (25, 8) {\small $1$}
            \put (47, 8) {\small $2$}
            \put (68, 8) {\small $3$}
            \put (90, 8) {\small $4$}
            \put (37, 2) {\small Frequency $\nu$ $\left[\mathrm{Hz} \right]$}
            \put (-4, 15) {\small $0$}
            \put (-9, 24) {\small $100$}
            \put (-9, 33) {\small $200$}
            \put (-9, 41) {\small $300$}
            \put (-9, 50) {\small $400$}
            \put (-9, 58) {\small $500$}
            \put (-9, 66) {\small $600$}
            \put (-9, 75) {\small $700$}
            \put (-9, 83) {\small $800$}
            \put (-16, 30) {\rotatebox{90}{\small Amplitute $\mathcal{A}$ $\left[ \cdot \right]$}}
        \end{overpic}} \qquad \quad
    \subfloat[Keta2]{\label{fig:Confronto_spettro_Keta2_sim}\hspace{0.5cm}
        \begin{overpic}[width=.4\textwidth]{Image_spectrum_Keta2.pdf}
            \put (3, 7) {\small $0.0$}
            \put (14, 7) {\small $0.5$}
            \put (25, 7) {\small $1.0$}
            \put (36, 7) {\small $1.5$}
            \put (48, 7) {\small $2.0$}
            \put (59, 7) {\small $2.5$}
            \put (70, 7) {\small $3.0$}
            \put (81, 7) {\small $3.5$}
            \put (93, 7) {\small $4.0$}
            \put (37, 1) {\small Frequency $\nu$ $\left[\mathrm{Hz} \right]$}
            \put (-4, 15) {\small $0$}
            \put (-9, 28) {\small $200$}
            \put (-9, 39) {\small $400$}
            \put (-9, 52) {\small $600$}
            \put (-9, 64) {\small $800$}
            \put (-11, 75) {\small $1000$}
            \put (-16, 30) {\rotatebox{90}{\small Amplitute $\mathcal{A}$ $\left[ \cdot \right]$}}
        \end{overpic}}
    \caption{\small Average frequency spectrum, comparison between the simulation \textbf{(a)} and the experimental data that have been used for the generation of the activation states in the simulation \textbf{(b)} (same plot as in Figure \ref{fig:Confronto_spettro_Keta2}).}
    \label{fig:Confronto_Spettro_Sim_Keta2}
\end{figure}
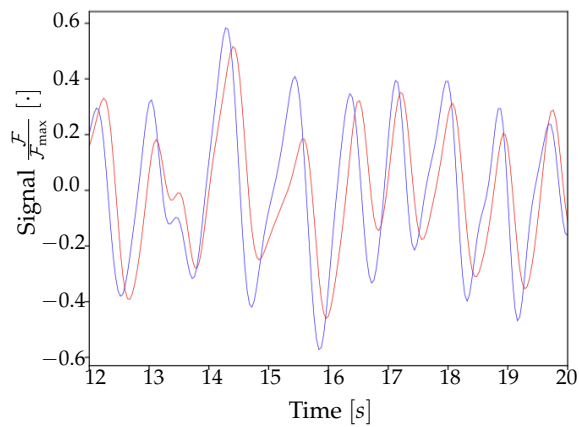
\begin{figure}[h!]
    \centering
    \captionsetup{skip=1cm}
    \hspace{0.75cm}
    \begin{overpic}[width=0.41\textwidth]{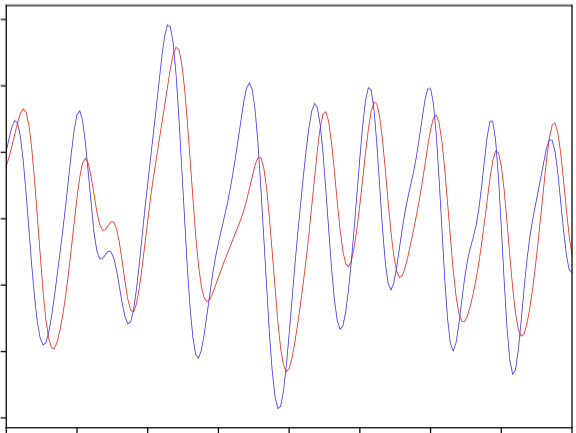}
        \put (0, -3) {\footnotesize $12$}
        \put (12, -3) {\footnotesize$13$}
        \put (24, -3) {\footnotesize $14$}
        \put (36, -3) {\footnotesize $15$}
        \put (49, -3) {\footnotesize $16$}
        \put (61, -3) {\footnotesize $17$}
        \put (73, -3) {\footnotesize$18$}
        \put (85, -3) {\footnotesize $19$}
        \put (97, -3) {\footnotesize$20$}
        \put (42, -10) {\small Time $\left[ s \right]$}
        \put (-9, 1) {\footnotesize $-0.6$}
        \put (-9, 13) {\footnotesize $-0.4$}
        \put (-9, 25) {\footnotesize $-0.2$}
        \put (-6.5, 37) {\footnotesize $0.0$}
        \put (-6.5, 48) {\footnotesize $0.2$}
        \put (-6.5, 59) {\footnotesize $0.4$}
        \put (-6.5, 71) {\footnotesize $0.6$}
        \put (-15, 28) {\rotatebox{90}{\small Signal $\frac{\mathcal{F}}{\mathcal{F}_{\mathrm{max}}}$ $\left[ \cdot \right]$}}
    \end{overpic}
    \caption{\label{fig:Confronto_segnale} \small Comparison between the signal generated by the simulation, in red, and the related experimental signal, in blue, for a single pixel.} 
\end{figure}

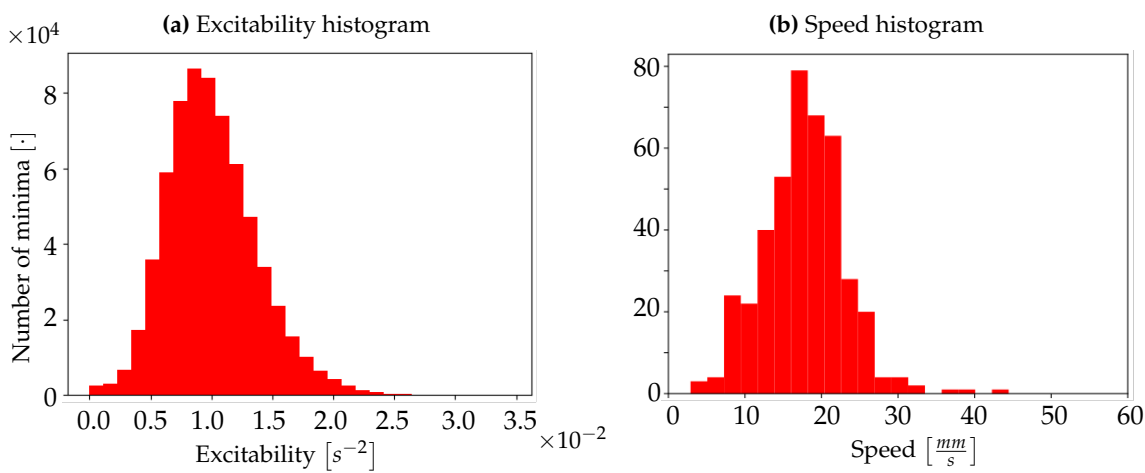
\begin{figure}[h!]
    \captionsetup[subfloat]{position=top}
    \hspace{1cm}\subfloat[Excitability histogram]{\label{fig:Sim_Results_histEcc}
        \begin{overpic}[width=.4\textwidth]{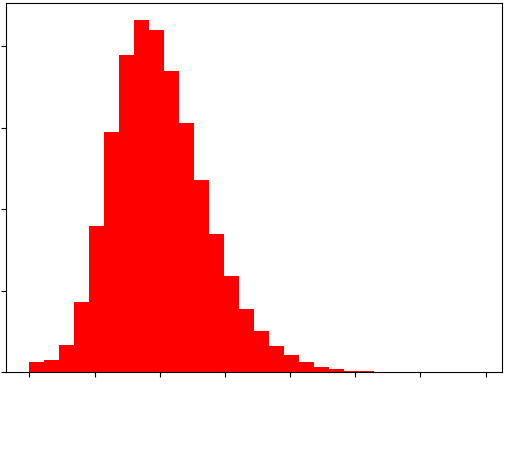}
            \put (3, 11) {$0.0$}
            \put (15, 11) {$0.5$}
            \put (27, 11) {$1.0$}
            \put (40, 11) {$1.5$}
            \put (53, 11) {$2.0$}
            \put (66, 11) {$2.5$}
            \put (80, 11) {$3.0$}
            \put (93, 11) {$3.5$}
            \put (100, 7) {\small $\times 10^{-2}$}
            \put (28, 4) {\small Excitability $\left[s^{-2} \right]$}
            \put (-4, 16) {$0$}
            \put (-4, 33) {$2$}
            \put (-4, 48) {$4$}	
            \put (-4, 64) {$6$}
            \put (-4, 80) {$8$}
            \put (-12, 93) {\small $\times 10^{4}$}
            \put (-11, 25) {\rotatebox{90}{\small Number of minima $\left[ \cdot \right]$}}
        \end{overpic}} \qquad \quad
    \subfloat[Speed histogram]{\label{fig:Sim_Results_histVel}\hspace{0.5cm}
        \begin{overpic}[width=.4\textwidth]{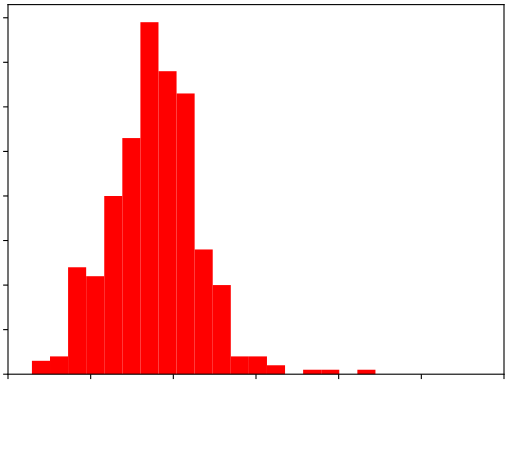}
            \put (1, 11) {$0$}
            \put (15, 11) {$10$}
            \put (32, 11) {$20$}
            \put (47, 11) {$30$}
            \put (63, 11) {$40$}
            \put (80, 11) {$50$}
            \put (96, 11) {$60$}
            \put (40, 4) {\small Speed $\left[\frac{mm}{s} \right]$}
            \put (-3, 16) {$0$}
            \put (-6, 34) {$20$}
            \put (-6, 51) {$40$}
            \put (-6, 68) {$60$}
            \put (-6, 85) {$80$}
        \end{overpic}}
    \caption{\small \textbf{(a)} Histogram of excitability for artificial data. \textbf{(b)} Histogram of wave propagation speed for artificial data.}
    \label{fig:SimHistos}
\end{figure}

Signals produced by the Toy Model are then submitted to the complete analysis pipeline, in order to obtain measurements of the average speed of wavefronts propagation and of the neuronal excitability. 
%
Comparing the histograms of propagation speed (Figure \ref{fig:WaveVelocity_Keta2} for Keta2 and Figure  \ref{fig:Sim_Results_histVel} for the simulation), we observe that the outcome of the Toy Model exhibits a different mean value, $\langle v \rangle = \left(18 \pm 6 \right)$ $\frac{mm}{s}$, resulting nonetheless compatible with the average speed of Keta2 within $1~\sigma$ error. This is indicative of the fact that the timing sequence of transitions, associated with the Lognormal transfer function, is sufficient to reasonably reproduce the experimental wavefront speed statistics.
Subtler differences in the distributions of speed cannot be easily attributed, since on the experimental side ketamine-induced slow rhythms present several irregularities, and on the simulation side the model is based on very simplistic assumptions.
Concerning the excitability, we do not include the excitability map of simulated data because, as expected, it does not present any spatial features; this is coherent with the assumption of the Toy Model in which \textit{Neurons} are intrinsically non-interacting, identical and highly stereotyped objects; the random pattern that appears in the excitability map reflects the randomness of the Poissonian processes, which are not bound to any spatial dependence. An effect of such randomness is also observable in the excitability histogram (Figure \ref{fig:Sim_Results_histEcc}), whose shape appears more Gaussian if compared to the distribution obtained with Keta2 data (Figure \ref{fig:Confronto_Ecc_hist_K2}), in agreement with the implication of the Central Limit Theorem, that results in a normal distribution after summing up a large number of independent Poissonian emitters.
The histogram should, in this sense, be interpreted as a validation of the intrinsic random character of the simulated model, rather than a reliable experimental measure of the values of excitability of the network.
%
Finally, the analysis of the origin points for the Toy Model has been omitted, as they are completely determined by the collection of transition times used as an input.

\section{Discussion}
\label{sec:Discussion}

In this paper, we discuss two different lines of work. The former consists in the analysis of experimental images: we introduced a complete sequence of analysis capable of identiying the slow waves signal on the cortex. The latter consists in framing the experimental signals in a simple theoretical model which takes into account the statistics of the neuron activity and the fluorescence response of the proteins encoded in the mouse brain. A simulator (denoted as \emph{Toy Model}) was implemented to synthesize artificial signals based on this model. Afterwards, the artificial signals were processed through the same pipeline developed for the experimental data, aiming at validating the analysis tools.

The first outcome we present is the release of the CaImanSWAP analysis pipeline. The preliminary version we distribute for the neuroscience community succeeds in providing a set of tools able to convert raw images (Calcium imaging data acquired with wide-field microscopy) in a format that can be received as an input by an already established analysis workflow whose developments have been guided by a completely different experimental set-up. The advantage of this approach is that the same scientific problem (in this case, the dynamics of the slow wave activity in deeply anaesthetised brains) can be faced enriching the statistics of data with differently acquired experimental samples, combining optical and multielectrode electrophysiology, and having the analysis tools as the unifying element that overcomes any systematics that could be due to the data acquisition. In this framework, quantitative comparisons of different datasets and of experiments and simulations can be enabled, without introducing any tailored adjustments in the signal processing. 
In agreement with objectives and conclusions stated in \citep{Afrashteh2017:OpticalFlowAnalysisToolbox}, we aim at going beyond visual inspection, enlarging the offer of tools that support quantitative studies of neuronal dynamics from imaging data; in addition, we pursue an effort towards the adoption of free software, potentially enlarging the number of users. 
Entering the details of other analysis solutions, approaches like the ``optical-flow'' generate velocity vector fields to determine speeds and directions of propagated activity (\eg in \citep{Afrashteh2017:OpticalFlowAnalysisToolbox, Townsend2018:DetectionAnalysisSpatiotemporalPatterns}); conversely, our tool is not associated to vector fields but infers the spatiotemporal propagation features based on the source signal (\ie localization of minimum values across the pixel field). Moreover, to estimate the major direction of propagation, the optical-flow approach relies on the selection of a region of interest, while our approach is not biased by this a priori choice. Finally, in the slow wave field, other studies reported analysis based on Granger causality (\eg in \citep{Mitra2018:SpontaneousInfra-slowBrainActivity}), with the limit of an arbitrary choice of the region of interest and of not being able to perform single wave propagation analysis. 

On the basis of what above commented, the first direction of development that we foresee is to adopt the tools here discussed to compare optical data and electrophysiological measurements. Our approach consists in having a common analysis pipeline that branches at the very beginning, when each type of experimental data shows its own specificity and requests a dedicated treatment for extracting the information from the raw signals. Once that the raw data are processed to get a collection of ``trigger'' times (\ie whatever could be interpreted as related to transition times or activation times) over a set of ``channels'' (\ie pixels or electrodes) with a defined geometry, such a collection can be set as an input for the common next steps of processing, regardless of the peculiarity of the raw data underneath. In particular, a crucial point of the pipeline is the WaveHunt algorithm, that splits the collection of times into a collection of waves. Here, beside improvements and further validation processes that could be included in the algorithm, it is where the wave analysis starts, that is the core of the quantitative investigation of the phenomenon. In addition to the measurements we have presented in this work, several other studies can be carried on, for instance: a principal component analysis (PCA), to identify the content of variability in the dataset and to define a parameter space where performing the wave clustering; the evaluation of quantitative measurements (such as the mean speed and the mean propagation pattern) for each identified cluster. Some preliminary tests are ongoing, aiming at a direct comparison of an extended set of experimental data, necessary for a solid statistical assessment of the findings and a larger significance of results.

A further direction of study could concentrate on data samples acquired from subjects in different neurophysiological states, but still manifesting slow wave activity; the approach we have developed could enable a quantitative comparison of different anesthetics (as GABAergics such as propofol, isoflurane and sevoflurane); also, the option of measuring the effects at different levels of anesthesia (from deep anesthesia towards wakefulness) could be evaluated.

The current status of the analysis pipeline is preliminary and there is large room for improvements, addressing different aspects of the processing. 
One of them is the filtering of signals. In the present stage, we adopted Butterworth filters, that are designed to have a frequency response as flat as possible in the passband (maximally flat magnitude filters), but introduce phase distortion that can affect the spatiotemporal patterns. 
As opposite, linear phase filtering is reached with different approaches, as Bessel filters (designed for a maximally linear phase response in the passband), filters with symmetric impulsive response as causal FIR (finite impulse response) filters, or by filtering the signals bidirectionally, to compensate any phase distortion and optimize for a constant group delay. 
Beside, the impact itself of the phase distortion should be evaluated; indeed, if a maximal impact is expected in audio applications for which a minimal distortion is required, the request could be less stringent for the kind of problem we are facing, since, assuming a comparable spectral content for the pixels, the same phase distortion is introduced in all the channels, and the determination of the timestamps could be not significantly affected. Conversely, the phase distortion issue should be taken into account if trying to correlate the pixels'signal with other inputs with different spectral content and sampling dynamics, as LFP (Local Field Potential) data.
%
In addition, the order of filtering should be investigated for optimising the performance of the analysis.

As already anticipated, actions can be implemented also in the WaveHunt algorithm, to improve the performance. Here, in particular, the focus is to define additional (enriched or complementary) criteria for extracting waves from the collection of trigger times that the algorithm receives as input. As pointed out in note \ref{note:WaveHunt}, the obtained collection of waves depends on the defined criteria, and the current implementation favours the inclusion of  planar  and spherical global waves (in the presented setup, the globality threshold is set to the participation of 75\% of channels), ruling out overlapping patterns, and implies the assumption of regular wavefronts as the dominant subset of propagation modes. This introduced limitation is partly justified by the results of the visual inspection of the raw data, for the movie obtained with the raw images shows a dominance of regular propagation across the entire portion of cortex under observation. Taking the Slow Wave Activity (SWA) regime as the state exhibiting a regular pattern of propagation of the spiking activity in opposition to the asynchronous regime (awake state), the definition itself of a proper ``wave'' can be restricted to regular propagation patterns, \ie patterns fulfilling the unicity request we adopt in the pipeline, thus excluding cycloidal or aborted propagation. In addition, a more quantitative evaluation of the different propagation patterns could be reached introducing complementary criteria, in order to quantify the relative weight of different criteria and the relative extent (cardinality) of the wave collection depending on the criteria. In other words, since some selections may be mutually exclusive\footnote{For instance, the unicity and the globality principles cannot be compatible with the isolation of overlapping partial waves.}, the approach should be to define different sets of selection criteria, identifying for each set the capabilities and the limitations, and comparing the resulting collection of waves. This approach should be taken into account as a general guideline in the analysis of data, regarding any step of the processing. Indeed, part of our speculations about the pipeline consists in identifying how results and outputs are dependent on the analysis settings, since the unavoidable manipulation that data analysis introduces has an impact and may enlighten/isolate a given feature of the data or hide some other. One of the ambitions we have with the pipeline named SWAP is to equip the data analysis procedures with tools that enable the control and monitoring of the applied analysis criteria, in order to keep track of the adopted choices and to identify compatible and incompatible options or dependencies, towards a solid description of the processing steps aimed at a reliable comparison of datasets.

Concerning the Toy Model, the comparison between experimental and simulated data produced a reasonable agreement; future extended analysis may help figure out additional features, for a deeper understanding of the phenomena. Nevertheless, despite the current simplicity of the model, our findings, obtained just by adopting a LogNormal transfer function inspired by the experimental response described in \citep{Chen2013:Ultrasensitive-fluorescent}, suggest that the spiking activity, among the many contributions, is one of the principal components of the observed fluorescence fluctuations. The analytical description of the transfer function could be further refined using the Toy Model, assuming a given description dependent on a set of parameters and training a decision algorithm, such as a neural network, to infer the parameters, that could eventually be adopted in performing a deconvolution process.

The reconstruction of the wavefronts obtained with this analysis can also be used for the estimation, with a high spatial resolution, of effective connectivity matrices 
between the neuronal populations underlying the individual pixels.
Eventually, the high-resolution spatio-temporal characterization of the Slow Wave Activity enabled by our methods could be an important ingredient to tune the parameters and the interconnect of spiking neural network simulations, like those empowered at similar spatio-temporal resolution by distributed engines \cite{Pastorelli2018:Gaussian-and-Exponential, Paolucci2018:Sleep-like-slow-oscillations}. This data-driven simulation approach will overcome the limits of current simulations that use simplified and stereotyped assumptions, and it will make possible the validation of competing theoretical models, by matching simulations with experimental data. 

\vspace{6pt} 



\authorcontributions{Conceptualization, F.R, A.L.A.M, F.S.P., G.D.B. and P.S.P.; methodology, all; software, M.C., C.D.L. and P.M.; investigation, F.R., A.L.A.M; writing -- original draft preparation, M.C., C.D.L. and P.M.; writing -- review and editing, F.R, A.L.A.M, G.D.B. and P.S.P.; visualization, M.C., C.D.L. and P.M.; supervision, P.S.P and F.S.P.; project administration, P.S.P and F.S.P.; funding acquisition, P.S.P and F.S.P.}

\funding{This research was funded by the European Union’s Horizon 2020 Framework Programme for Research and Innovation under \hbpgrant.}

\acknowledgments{This study was carried out in the framework of the Human Brain Project (HBP\footnote{\url{https://www.humanbrainproject.eu/en/}}), funded under \hbpgrant, in particular within activities of sub-project SP3 (``Systems and Cognitive Neuroscience'') and sub-project SP1 (``Mouse Brain Organization''). Part of this work has been submitted in fulfillment of the requirements for the final exam of the course of Physics Laboratory II (Biosystems) 2017--2018, held by Prof. Federico Bordi for the Master's Degree in Physics, ``Sapienza'' University of Rome.}

\conflictsofinterest{The authors declare no conflict of interest.}

\reftitle{References}

\externalbibliography{yes}
\bibliography{INFN-LENS.bib}



\end{document}